# Perceptual Requirements for Low-Latency Head-Mounted Displays

ERIC PENNER, Reality Labs Research, Meta, USA
JOSEPHINE D'ANGELO, Reality Labs Research, Meta University of California, Berkeley, USA
CLINTON SMITH, Reality Labs Research, Meta, USA
NATHAN MATSUDA, Reality Labs Research, Meta, USA
NEETHAN SIVA, Reality Labs, Meta, USA
PHILLIP GUAN, Reality Labs Research, Meta, USA

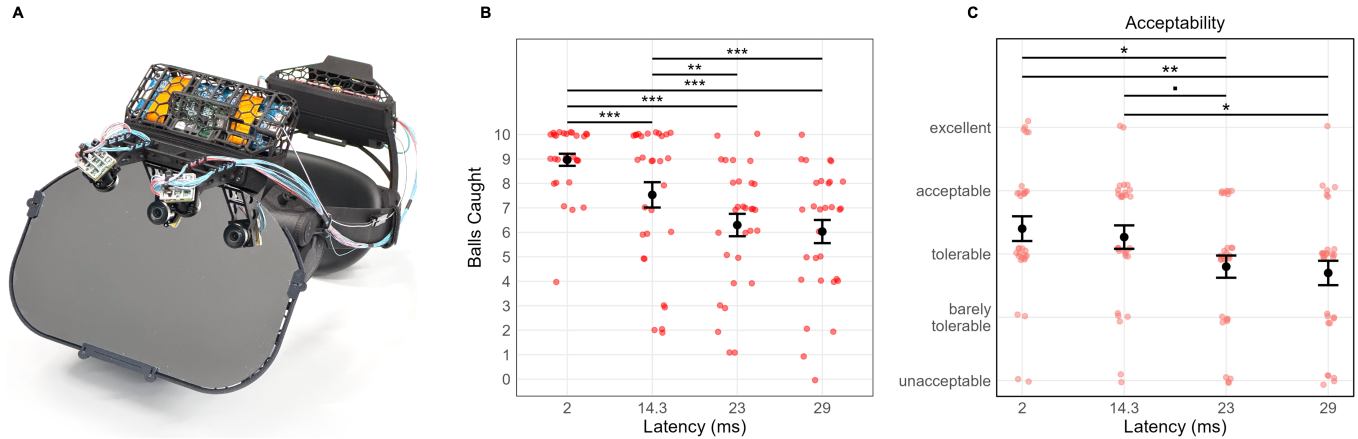

Fig. 1. **(A)** We present a video passthrough headset achieving 2 millisecond end-to-end (e2e) latency. A catadioptric design employs world-facing mirrors to align the passthrough cameras with the user's physical eye position in the HMD, minimizing perspective error to ensure latency is the primary variable under study. **(B)** Across two studies (N=27, N=30) we evaluated the impact of e2e latency on task performance and subjective evaluations between 2 and 29 milliseconds. Here we show participant ball-catching performance from the second study. Performance is best at two milliseconds and declines with increasing latency. **(C)** Subjective responses from the second study also reveal a significant preference for 2 and 14.3 milliseconds of e2e latency over 23 and 29. In our second study we additionally collect psychophysical measurements of latency detection thresholds in a non-headset form factor testbed to explore how psychophysical thresholds can be paired with naturalistic studies user experience evaluations.

End-to-end (e2e) latency in head-mounted displays (HMD) is the time delay between a physical change in the world (e.g., a user's head movement) and the moment the display updates to reflect that change. Tracking, rendering, and other computation in real systems invariably introduce some amount of e2e latency to all HMDs. In modern devices this latency is usually in the range of 12-60 milliseconds which is partially addressed through pose prediction and late stage reprojection which means that perceptual studies and user experience evaluations cannot explore latencies below these values. Here, we introduce a video passthrough HMD, called Camsicle, which is capable of 2-millisecond e2e latency and, additionally, uses a catadioptric design to achieve perspective-correct passthrough without reprojection. This platform enables naturalistic user studies to interrogate the impacts of latency on user experience, preference, and performance. Across two user studies and 57 participants we find that 2 and 14.3 millisecond latencies are preferred over 23 and 29 milliseconds when attempting to catch a ball. Additionally, we compare individual latency preferences in this naturalistic ball-catching task to psychophysical thresholds for latency detection in a reference-grade system with zero latency to investigate how psychophysical thresholds may relate to subjective evaluations in naturalistic scenarios.

## 1 INTRODUCTION

Egocentric, head-mounted displays (HMDs) create immersive 3D environments by rendering and presenting perspective-correct stereo images that respond dynamically to the viewer's motion. However, the efficacy of this illusion is fundamentally constrained by the system's ability to minimize errors in stereo geometry. Errors in perspective projection—whether caused by tracking inaccuracies [Holloway 1997], latency [Allison et al. 2001], or optical distortions from near-eye lenses [Geng et al. 2018; Tovar et al. 2024]—can introduce visual-vestibular cue conflicts. These conflicts degrade immersion and are primary drivers of visually induced motion sickness (VIMS).

End-to-end (e2e) latency represents the delay between the recording of a physical movement or change in the world and the corresponding update of photons on the display (we use e2e because it encompasses both motion-to-photon and photon-to-photon latency). The presence of e2e latency means that stereo images seen by the viewer are not perspective correct, and the magnitude of those errors depends on the total latency and user or world motion. Any processing time or delays in the entire HMD pipeline—including tracking, rendering, and display presentation—contribute to a system's total e2e latency. Minimizing e2e latency is particularly important for augmented reality (AR) and video passthrough systems. In AR, the physical world serves as an instantaneous, zero-latency reference, and any temporal lag in the HMD causes virtual content to visually "swim" or drift relative to the real world, which makes inconsistencies more easily perceptible [Mckee and Nakayama 1984]. In passthrough headsets, the entire passthrough feed is delayed so the world may appear world-locked and stable, but haptic and proprioceptive feedback from physical interactions with the world will not be aligned with visual input.

Commercially available headsets typically operate with latencies in the range of 30–60 milliseconds with one notable exception being the Vision Pro which has a reported latency as low as 12 milliseconds [Deveci and Sarviluoma 2025; Gruen et al. 2020]. VR headsets necessarily introduce latency during frame rendering, and HMDs employ compensatory strategies such as forward prediction, asynchronous time warp, and view reprojection [Azuma and Bishop 1994; Freiwald et al. 2018; Rong 2025; Van Waveren 2016; Warburton et al. 2023]. While these techniques effectively reduce the perceptual impact of latency, they introduce complexity for research: it is difficult to study the implications of latency on hardware employing these techniques because the absolute latency and efficacy of these compensations is difficult to quantify.

Passthrough video bypasses the traditional 3D rendering pipeline, and its photon-to-photon latency is theoretically limited by camera exposure time rather than scene complexity. However, conventional implementations of video passthrough introduce an additional perspective offset between the user's eye position and the passthrough camera location on the device. Leaving this offset uncorrected introduces stereo geometry errors, but correcting it—whether through view reprojection [El Chemaly et al. 2025; Freiwald et al. 2018; Xiao et al. 2022] or computational imaging techniques [Kuo et al. 2023]—introduces potential software confounds and artifacts. Custom hardware and rendering software has been built with extremely low latency [Regan et al. 1999], but such low-latency research prototypes often lack the form factor or portability required for naturalistic tasks [Guan et al. 2023] and limit the ecological validity of stimuli available for research.

In this work we combine both naturalistic and psychophysical study designs to comprehensively evaluate e2e latency in HMDs. In doing so we make the following contributions:

- We present Camsicle, a custom-built video passthrough headset achieving an e2e latency of just 2 ms. The device utilizes a catadioptric optical design to physically align the passthrough cameras with the viewer's eye position, minimizing geometric errors from inaccurate parallax thereby eliminating need for view reprojection.
- Using Camsicle, we conduct a naturalistic evaluation (N=27) of e2e latency on three distinct tasks (ball-catching, a reaction time measurement, and a timed pen and paper maze) at 2, 14.3, and 29 ms. We identify declines in objective and subjective measures as latency increases these three values.
- In a follow-up study (N=30) we explore if an individual's psychophysical detection thresholds can predict individual differences in naturalistic evaluations of latency. We do not find any statistically significant relationships, but highlight potential modifications to the psychophysical experiment for future work.

## 2 RELATED WORK

*Latency Geometry.* In the context of HMDs, latency manifests as errors in stereo projection geometry [Azuma 1997]. Holloway [Holloway 1997] categorized these errors into static (viewing/rendering offsets) and dynamic categories. They demonstrated that for typical head rotation speeds, system latency (at the time) can be the single largest contributor to what they called geometric registration error. Other works have built models of stereo geometry for spatial viewing or rendering errors that do not explicitly account for latency, but the frameworks can be easily modified to account for spatial errors from latency as well [Held and Banks 2008; Rolland et al. 2004; Wann et al. 1995; Woods et al. 1993]. Jerald [2010] also use a model to generalize from head motion to define a perceptual limit for allowable latency, and propose a limit of 5 ms.

*User Impacts of Latency.* Latency has been shown to degrade performance in remote manipulator control [Sheridan and Ferrell 1963], target acquisition [MacKenzie and Ware 1993], reaching and selecting objects [Ware and Balakrishnan 1994], and placing objects in 3D environments [Watson et al. 2003]. Physiologically, excessive latency can be a driver of simulator sickness [DiZio and Lackner 2000; Palmisano et al. 2017, 2024; Stauffert et al. 2020]. Others have shown found that higher latency reduces vection [Kim et al. 2022] and feelings of presence [Welch et al. 1996]. Perceptually, this can also manifest as a loss of world-stability where the visual world appears to oscillate, a phenomenon known as Oscillopsia [Allison et al. 2001; Bender 1965]. Elbamby et al. [2018] cite "broad census" that e2e latency should kept below 15–20 ms to provide a pleasant immersive experience, but such numbers are difficult to validate given the black box nature of a typical HMD tracking, rendering, and presentation pipeline. Our Camsicle headset addresses these concerns through its simple and straightforward software pipeline. The passthrough nature of our system ensures that visual stimuli are well-defined and the low latency-floor ensures that naturalistic evaluations of latency can be done without confounds from the baseline latency of the HMD itself.

*Psychophysical Measures of Latency.* Psychophysical paradigms have previously been used to identify latency detection thresholds. Regan et al. [1999] build a mechanically-tracked stereoscopic "fish tank" VR system with only 200 microseconds of e2e latency and reported latency detection thresholds between 5-25 ms across 12 participants. More typical hardware used by others to measure latency have much higher baseline latency which fundamentally limits the lowest latencies that can be studied. For example, Allison et al. [2001] found that 50% latency thresholds for oscillopsia ranged from 180–320 ms when added to their system's baseline latency of 120 ms. These thresholds would likely be lower if they had employed a two-alternative forced choice paradigm instead of a yes/no design. Others have also reported latency detection thresholds in the range of 5-20 ms, but these results are derived using non-standard definitions for a "just noticeable difference" (JND) applied to atypical psychometric functions showing detection rate rather than percent correct. These data are used to find what the authors call the "point of subjective equality," but is typically defined as the fist JND [Adelstein et al. 2003; Ellis et al. 1999; Jerald and Whitton 2009]. We also use a mechanically tracked system to derive psychophysical latency detection thresholds. Unlike past work, our system incorporates accurate 3D eye position in rendering through the use of a calibrated bite bar. This enhanced accuracy allows us to accurately render perspective-correct stimuli and allows us to render virtual content locked to the real world, allowing us to measure latency detection thresholds for AR in addition to VR.

*Low-Latency Video Transmission.* Outside of AR/VR, video and display system latencies have significantly increased over the last 20 years. The earliest video broadcast systems used vacuum tube cameras to convert light into an analog signal, which was transmitted at near the speed of light until being received by a similar vacuum tube in a CRT display [Fink 1940; Zworykin 1934]. While most video mediums are not highly latency sensitive, a few have continued to optimize for latency, such as first-person-view (FPV) drone operation [Pfeiffer et al. 2021]. While some of these systems claim to achieve latencies similar to Camsicle [Divimath, Inc. 2024], these mono camera systems do not account for capture or viewing optics. Additionally, they are completely decoupled from the user so they also do not provide perspective-correct parallax required for AR/VR immersion.

## 3 LOW-LATENCY PASSTHROUGH

In order to investigate the impact of passthrough latency in a naturalistic setting, we built a headset capable of significantly lower latency than existing commercial products while retaining as much of the capabilities and form factor of such devices. A significant challenge in isolating the effects of latency is that most headset architectures utilize software latency-mitigation and perspective-correction schemes (such as timewarp and reprojection warping), which introduce additional perceptual effects on top of the theoretical minimum latency for the underlying optoelectronic system. However, we can eliminate these software subsystems if we build a headset that minimizes both perspective mismatch and latency simultaneously.

### 3.1 Hardware

We cannot directly eliminate perspective offsets as this would require co-locating cameras with the user's eyes. We also must avoid optical architectures that introduce additional compute costs (and thus latency) like the one proposed in [Kuo et al. 2023]. Instead, we employ a folded optical path using a lightweight mirror to approximately co-locate the passthrough camera's entrance pupil with the user's nominal pupil position in the headset (see Figure 2). To prevent the camera hardware from occluding the user's view, we incline the mirror which shifts the reflected camera assembly out of the FOV while maintaining optical co-location. To improve the size and weight, we cut the mirror using a rounded FOV mask matched to the headset's display FOV. The resulting mirror weighs 62.5 grams for a total headset weight of 705 grams, which is on par with contemporary commercial headsets. Furthermore, we select camera and display optics that have approximately matched distortion profiles, to minimize the distortion correction software requirements.

As discussed in section 1, latency in a passthrough system can be measured as the time difference between recording a stimulus and presenting it. For finite, discrete sensor and display integration times, we can consider the time differene between the midpoints of the camera integration window, and the displays emission period (set by the display persistence in a head-mounted display). A conventional passthrough pipeline first accumulates a framebuffer for the outward-facing cameras. Once this is transmitted to the headset's compute subsystem, a series of image processing operations are applied, including distortion and lateral color correction for the passthrough camera lenses, compositing with virtual content, the aforementioned software latency correction, and distortion/lateral color correction for the display optics. Then the entire framebuffer is transmitted to the display subsystem, and it is subsequently presented to the user. This process takes, at a minimum, the combination of the frame time of the camera plus the frame time of the display, and in practice results tens of milliseconds latency.

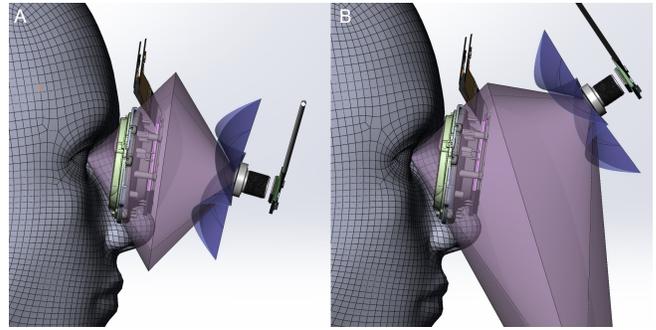

Fig. 2. **(A)** Folding the optical path of the camera with a mirror places the camera's entrance pupil at the user's nominal pupil position, resulting in perspective correct passthrough. However, the camera looks directly back at itself, blocking the most important part of the user's FOV. **(B)** By placing the mirror at an angle, the camera moves upward out of the FOV. As the mirror tilt increases, the mirror must grow in size, so a trade-off must be made between mirror size and partial camera occlusion when supporting large FOVs.

The CMOS sensor and OLED display components in modern VR headsets, however, operate in a row-by-row manner, with full-frame buffering being an additional stage implemented in camera drivers and graphics pipelines to support the convention of frames as the atomic unit for video sequences. We reach back to older techniques designed for cathode ray tube displays that "race the beam" [Lincoln et al. 2016; Regan et al. 1999], eliminating the delays associated with accumulating entire images.

We cannot operate purely on a per-row basis as the optical distortion correction steps result in display rows becoming curved rows in camera space. All required camera pixels for a display row must have arrived before a display row is composited. We can, however, approximately match the distortion profiles and scan timings of both camera and display, such that a minimal number of camera rows must be buffered (see Figure 4), with the primary drivers of latency being the number of rows in this camera buffer, and the amount of lead time "ahead of the beam" we trigger our rendering jobs.

We implement this approach using an FPGA camera driver that streams row-data into a locked shared memory buffer. A custom compositor performs the entire camera processing pipeline (e.g., demosaicing, color correction, and distortion) in a streaming fashion, writing the results directly to the GPU's front buffer. Bypassing the conventional frame-based architecture in this way allows us to reduce e2e latency by an order of magnitude. Please see our supplementary materials for implementation details.

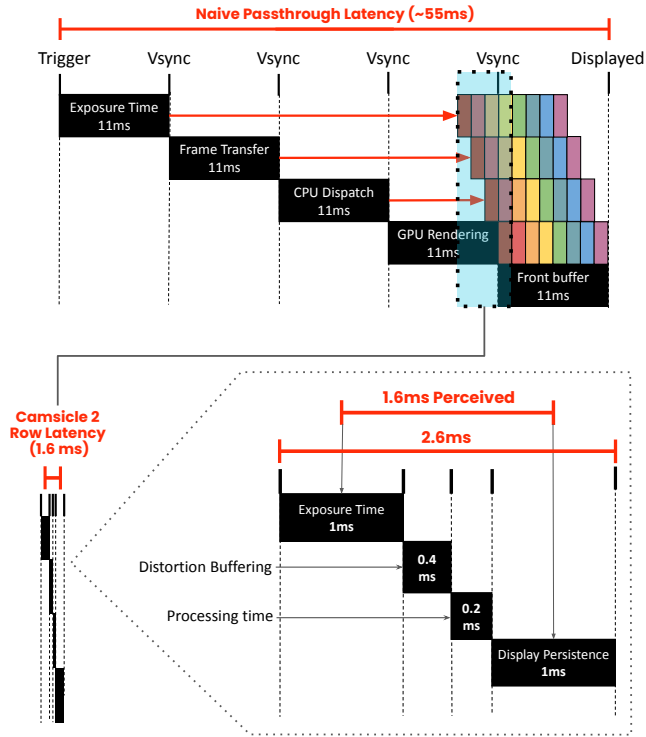

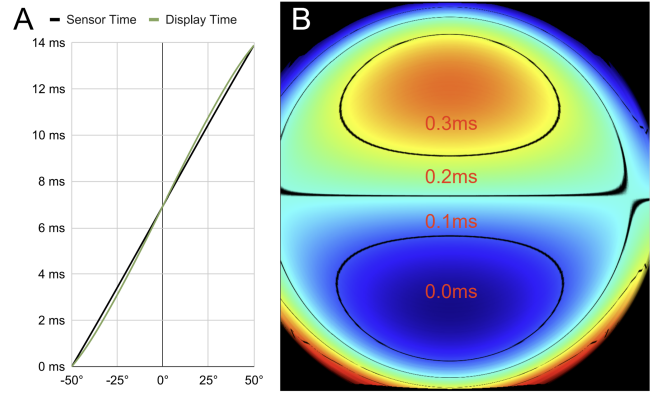

Fig. 3. Top: Comparison of a naive passthrough pipeline to Camsicle. By precisely synchronizing a rolling shutter camera and display, and slicing the frame into 8 slices, an almost 8-fold improvement in pipeline latency can be realized. Bottom: By taking this to the limit with co-designed optical paths, our system is capable of reaching as low as 1.6 ms of perceived latency, of which only 0.2 ms comes from the software pipeline.

### 3.2 Software Pipeline

### 3.3 Calibration and Measurement

We perform initial camera-display calibration using a through-the-lens witness camera following [Kuo et al. 2023], and refine using ChArUco board calibration [Garrido-Jurado et al. 2014] to account for minor deviations from our simulated lens distortion.

We embed timestamps in each row of our camera data to measure the latency of the software components. Timestamps are embedded immediately after camera exposure is completed and read by the FPGA. The latency for a rendered pixel can calculated from source camera row's timestamp, the target row's display time, and the vsync clock. Since the graphics driver or operating system could in theory provide a false view of our display's latency, we also verified both our display's latency and end-to-end latency with a 1920fps high speed camera. Our lowest stable latency was 1.6 milliseconds, when rendering 0.2 milliseconds ahead of the beam, with 0.4 milliseconds of camera buffering (see Figure 3).

Fig. 4. **(A)** Simulated sensor and display times for the center column of the camera and display, which match very closely in our final optical configuration. Times are calculated by applying the simulated distortion curves followed by calculating the time for that display/camera row. **(B)** The difference between the two curves in 2D, in final calibrated display space, with a phase offset to avoid negative latencies. The heatmap plots the resulting latency we can achieve for each pixel, which is uniformly below 0.4 milliseconds, and dictates the amount of camera buffering required.

### 3.4 Camsicle Fitment

The display and passthrough camera separations are both fixed at 64 mm. This fixed design is a compromise to increase device rigidity (which improves direct passthrough accuracy) and reduces weight and complexity. Prior work has shown that, perceptually, matching participant interpupillary distance with display and "render" camera separation may not be as important as correctly matching eye-relief values [Allison and Wilcox 2015; Zhu et al. 2025]. We designed a custom facial interface with magnetic spacers that allowed us to modify the viewer's eye position in the headset by one millimeter increments. A magnetic attachment with a rear and front boresight was used to align the viewer's corneal apex to the intended eye relief (Figure 5).

## 4 LATENCY IN NATURALISTIC TASKS

In our first set of user studies, we examined how different levels of latency influence user sentiment and task performance, as well as participants' ability to distinguish between latency conditions using forced-choice ranking. Participants engaged in three tasks—ball catching, a timed maze, and reaction time measurement—while wearing our the low-latency video passthrough headset described in Section 3 configured to present latency values of 2, 14.3, and 29 milliseconds while participants completed each of the three tasks.

### 4.1 Participants

The study was conducted in accordance with institutional ethics and safety standards. A total of 27 individuals participated in the two-hour study (17 female, 8 male, 2 did not respond), with ages ranging from 18 to 74. The age distribution was as follows: 11 participants

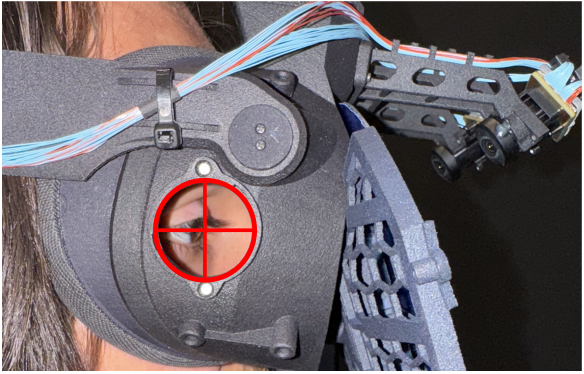

Fig. 5. Facial interface with cover removed for boresight alignment. A virtual boresight overlay is shown here in place of the actual attachment to demonstrate the alignment principle. One millimeter thick magnetic shims can be added or removed from the facial interface for depth adjustment to ensure precise eye relief for each participant.

were 18–24, 6 were 25–34, 4 were 35–44, 1 was 45–54, 2 were 55–64, and 3 were 65–74.

### 4.2 Methods

The order of latency conditions and task presentation was counterbalanced across participants to minimize order effects and learning biases. The actual latency value was blinded to each participant, and they completed three experimental blocks corresponding to each latency condition. Counterbalancing was done first on e2e latency value, and within each latency block the order of the three tasks was counterbalanced as well. Participants were blinded to the latency condition in all blocks to prevent expectation bias.

After each task was completed within a latency block participants rated their own subjective performance (1 = failure, 5 = perfect), perceived effort (1 = very low, 5 = very high), and frustration (1 = very low, 5 = very high) using a five-point Likert scale. Upon completion of all three tasks in a latency block, participants additionally rated the overall acceptability (1 = unacceptable, 5 = excellent) and responsiveness (1 = very poor, 5 = very good) of all three experiences collectively. Participants were permitted to adjust their ratings from previous sessions after completing each task, ensuring that their relative subjective assessments reflected their most current impressions. Each of these five surveys and their specific phrasing can be found in supplementary materials. The entire session for each participant lasted approximately two hours, including headset fitting, baseline practice of each task without the headset, and additional free form responses after the all three blocks were completed.

*Ball Catching.* In the ball catch task, participants attempted to catch 10 balls, and we used an automated pickleball trainer (Simon X Pickelball Machine) to throw each ball so that ball travel profiles were similar across trials. The trajectory of each ball was set to mimic a gentle underhand toss, and participants were asked to reach out and catch the ball with one hand using an overhand grip to maximize the precision required to successfully catch each ball. We found that this approach highlighted the effects of latency by making differences in an individual's anticipated ball trajectory and the actual ball location more apparent compared to two-handed and underhanded catching. Performance in ball-catching was defined by the number of successful catches.

*Timed Maze.* Participants completed a printed maze with only one valid path, and no route planning or higher level strategy needed to complete it. During the no-headset baseline training, participants completed the maze as quickly as possible while aiming to avoid touching maze boundaries to set a individualized baseline time target for each of the three headset latency conditions. Performance in this task was defined as the average number of intersections with maze boundaries, and three mazes were completed in each latency block. Participants were given general instructions but were not informed of the specific scoring criteria.

*Reaction Time.* We built a reaction time measurement box modeled after an open-source Arduino project[1]. In our device an LED is dimmed once a participant depresses a button, and they are instructed to release the button once the LED turns back on (sometime between 2-7 seconds after the button is pressed). A display on the back recorded each participant's reaction time, and five measurements were taken in each latency block. Performance is defined by each participant's reaction time; we did not show participant's their reaction times during the experiment, but did reveal these numbers to them after the entire two-hour experiment was over if requested.

### 4.3 Statistical Analysis

We use general linear mixed models (GLMM) to evaluate the impacts of latency on performance and Likert ratings in our three tasks. The specific model used depends on the outcome types (Likert rating, binary catch success, cumulative count of mistakes, or continuous reaction time). For each GLMM, latency can either be treated as a continuous, ordered categorical, or unordered categorical variable. In our analysis, we treat latency as an unordered categorical variable to enable direct pairwise comparisons between the three latency conditions used in the study. However, this approach does not model any potential ordinal or continuous relationship between latency levels, and thus does not capture trends or gradients that may exist across the latency settings. By ignoring the inherent relationship between latency conditions, this choice may decrease statistical power and sensitivity, but enables direct pairwise comparisons between the three conditions. To account for the increased risk of Type I error when performing these multiple comparisons, we also apply a Tukey p-value adjustment [Hothorn et al. 2008; Lenth 2023; Tukey 1949].

Every GLMM included a participant-specific random intercept to account for individual differences in baseline performance. Latency is treated as a fixed effect, which estimates a global coefficient representing the population-wide impact of latency deviations from those individual baselines. This approach leverages our repeated-measures design by accounting for individual differences in model fitting and modeling changes in an individual's performance across latency levels.

---

[1]https://projecthub.arduino.cc/rowan07/a-simple-reflex-game-dcd2c8

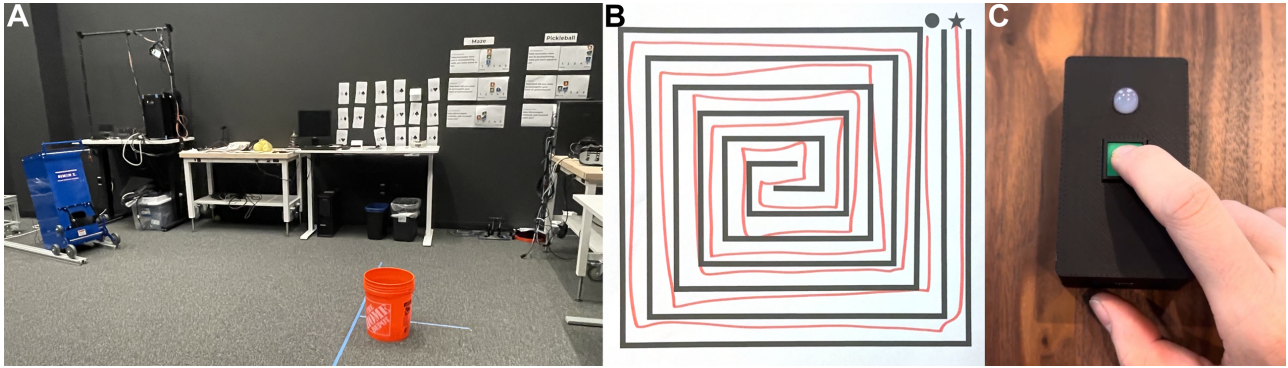

Fig. 6. Tasks used in our naturalistic evaluation of latency at 2, 14.3, and 29 milliseconds. **(A)** Catching a ball from a pickleball machine with a ball trajectory that matches the speed and arc of a gentle underhand toss. **(B)** Finishing a maze as quickly as possible without drawing over the maze boundaries. **(C)** Releasing a button when the LED turns on to measure reaction time.

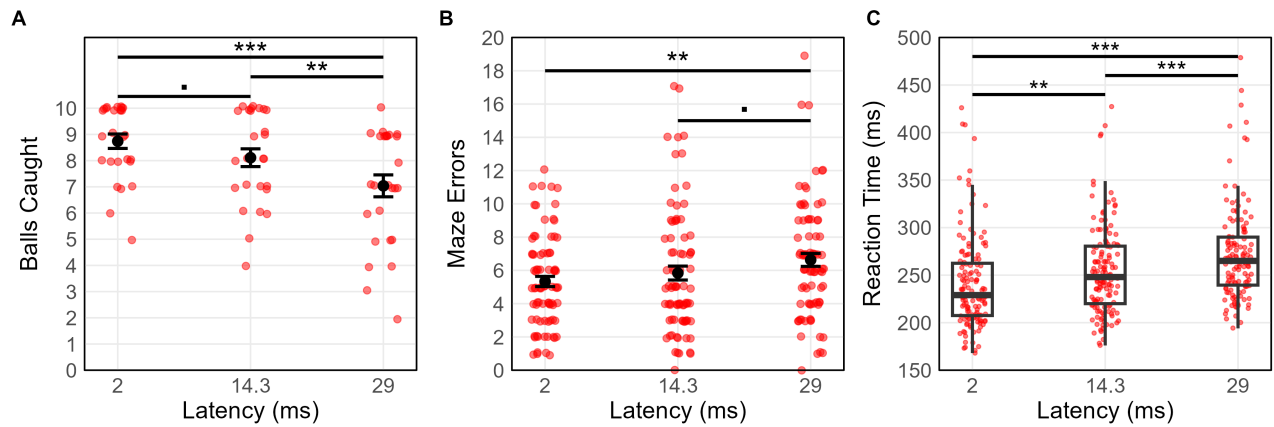

Fig. 7. Performance across tasks generally declines with increasing e2e latency. **(A)** Participants caught fewer balls as latency increased; pairwise comparisons show significant differences for both 2 ms and 14.3 ms compared to 29 ms, with a marginal trend between 2 ms and 14.3 ms ($p = 0.08$). **(B)** Participants drew over maze boundaries more often at 29 ms compared to 2 ms, while a marginal trend was observed between 14.3 ms and 29 ms ($p = 0.09$). **(C)** Participant reaction time showed statistically significant increases across all three latency conditions. Note: All $p$-values reflect Tukey-adjusted pairwise comparisons; ** $p < 0.01$, *** $p < 0.001$.

### 4.4 Task Performance

Participant performance results across all three tasks are summarized in Figure 7. Performance was significantly improved at 2 ms compared to 29 ms across all tasks. Similar significant improvements were observed when comparing 14.3 ms to 29 ms for the ball catching and reaction time tasks, with the maze task showing a similar trend ($p = 0.09$). Furthermore, the 2 ms condition significantly outperformed the 14.3 ms condition in the reaction time task, with a marginal improvement observed in ball catching ($p = 0.08$, after p-value adjustment for multiple pairwise comparisons). Consistent with these directional differences, the effect of latency is also statistically significant across all tasks when modeled as a continuous or ordered variable.

*Ball Catching.* Catch performance was analyzed using a GLMM with a binomial error distribution and a logit link function, including both fixed effects for latency and random intercepts for each participant (PID). The form of the model is specified as:

$$\text{Catch} \sim \text{Latency} + (1 \mid \text{PID}) \tag{1}$$

*Timed Maze.* Maze performance was analyzed using a GLMM with a Poisson error distribution and a log link function to model the count of total errors. The model included fixed effects for experimental condition and time spent to complete each maze, as well as random intercepts for each participant (PID). Unlike the other tasks, the maze navigation included an additional factor to account for the amount of time spent on each maze (i.e., if the same individual spent more time on the same maze we expect fewer mistakes). The form of the model is specified as:

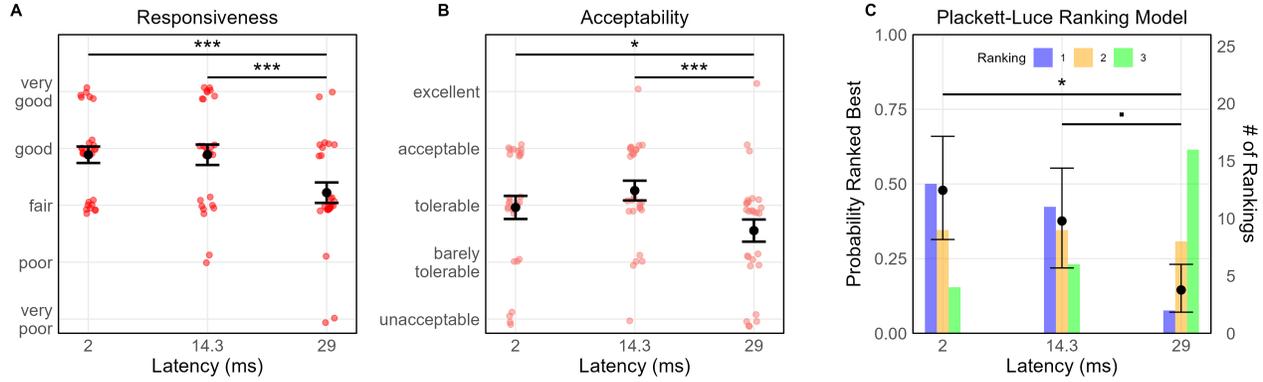

Fig. 8. Ratings for responsiveness and acceptability were collected at the end of each latency block, after all three tasks were completed. **(A)** Statistically significant changes in subjective impressions of responsiveness were found when comparing ratings at 2 and 14.3 ms to 29 ms ($p < .001$). **(B)** Statistically significant changes in acceptability rating were also found at 2 and 14.3 ms compared to 29 ms ($p = 0.03$, $p < 0.001$ respectively). **(C)** After all three latency conditions were completed (i.e., performing all three tasks a total of three times each, once for each latency condition) participants ranked each latency condition from best to worst. Participants were significantly more likely to rank 2 ms as the best condition compared to 29 ms ($p = 0.011$). A similar marginal trend was observed for 14.3 ms compared to 29 ms ($p = 0.053$).

$$\text{Total Errors} \sim \text{Latency} + \text{Time} + (1 \mid \text{PID}) \quad (2)$$

*Reaction Time.* Reaction time performance was analyzed using a GLMM with a Gamma error distribution and a log link function to model the positively skewed reaction time data. The model included a fixed effect for experimental latency (in seconds), and random intercepts for each participant (PID) to account for repeated measures within subjects. The form of the model is specified as:

$$\text{RT} \sim \text{Latency} + (1 \mid \text{PID}) \quad (3)$$

### 4.5 Subjective Evaluations and Rankings

Figure 8 shows participant ratings for responsiveness and acceptability, which were collected after participants completed all three tasks (totaling approximately 20 minutes of experience per latency condition). Task-specific survey results immediately after each task on self-rated performance, frustration, and effort are provided in the supplementary materials. Because the rating data are ordinal, results were modeled using a Cumulative Link Mixed Model (CLMM) with a Laplace approximation. The model included latency as a fixed effect and participant ID (PID) as a random intercept to account for repeated measures:

$$\begin{aligned}\text{Responsiveness} &\sim \text{Latency} + (1 \mid \text{PID}) \\ \text{Acceptability} &\sim \text{Latency} + (1 \mid \text{PID})\end{aligned} \quad (4)$$

Post-hoc pairwise comparisons with Tukey adjustment confirmed that both 2 ms and 14.3 ms conditions resulted in significantly higher ratings for responsiveness and acceptability compared to the 29 ms condition.

Participant forced-choice rankings were analyzed using the Plackett-Luce model, which is specifically designed for ranking data. This approach estimates the relative preference for each condition based on the observed rankings across all participants. The Plackett-Luce model was chosen over a CLMM because rankings are not independent judgments (i.e., if one latency is ranked as 1, the remaining latencies must be rated as 2 and 3). The model results indicated that the rankings for the 2 ms is preferred over 29 ms ($p = 0.01$) while the preference for 14.3 ms over 29 ms is marginally significant ($p = 0.053$). Taken together, these findings highlight a divergence between preference and performance; participants rated and ranked the 2 ms and 14.3 ms conditions similarly, yet the 2 ms condition afforded superior performance in the three tasks. Although the performance advantage of 2 ms over 14.3 ms was marginal in the ball catching task ($p = 0.08$) and not significant in the maze task, the directional trend is consistent across all metrics. This interpretation is further supported by the fact that latency remains a statistically significant predictor of performance in all tasks when modeled as a continuous or ordered variable (see supplementary materials for details).

## 5 RELATING PSYCHOPHYSICS TO USER EXPERIENCE

Naturalistic studies that focus on user experience during realistic tasks are essential for informing headset design and specification setting. However, these studies are challenging to scale, as subjective user experience is inherently ill-defined and responses are strongly influenced by individual differences. While within-subjects paradigms with repeated measures can improve signal-to-noise ratio and partially account for individual variance, such designs still require large sample sizes to achieve robust results. In contrast, forced-choice psychophysical methods for measuring latency detectability provides objective data with less variance since each threshold is derived from many trials instead of a single response. In this next study we present a study designed to explore the relationship between a participant's psychophysical latency detection threshold and their subjective evaluations of latency.

## 5.1 Participants

A new set of 30 participants participated in this study (15 female, 13 male, 2 did not respond), with ages ranging from 18 to 55. The age distribution was as follows: 9 participants were 18–25, 11 were 26–35, 6 were 36–45, 2 were 46–55, and 2 did not respond. We screened participants for ≤ 20/30 visual acuity using a Snellen eye chart, and for ≤ 40 arc seconds stereoacuity using a Randot test (Stereo Optical, Chicago, IL, USA). All experimental protocols were IRB approved.

## 5.2 Ball Catching Task

Based on free form feedback and ratings of subjective performance, effort, and frustration completed after each task, we decided to only use ball catching for the naturalistic task. The task was completed using the same protocol described in Section 4.2, with the same latency conditions of 2, 14.3, and 29 ms, and the addition of a fourth latency at 23 ms. We applied the same generalized linear mixed model (GLMM) analysis to assess both subjective and performance-based differences across latency conditions.

*5.2.1 Results.* Ball catching performance, subjective ratings, and ranking results are presented in Figures 1 and 9. Consistent with the prior study, we observe a divergence between subjective preference and objective task performance. While performance is best at 2 ms, subjective preference appears to function as a step, clustering into a "high preference" group (2 and 14.3 ms) and a "low preference" group (23 and 29 ms).

*Subjective Ratings.* Participants rated the acceptability and responsiveness of the system after each block. As shown in Figure 9, distinctions in Likert ratings were less sharp than in the ranking data. For Acceptability (Figure 1C), Tukey-adjusted pairwise comparisons revealed that 2 ms condition was rated significantly higher than both 23 ms ($p = 0.01$) and 29 ms ($p = 0.002$). The 14.3 ms condition was also rated significantly higher than 29 ms ($p = 0.01$), though the comparison against 23 ms only showed a marginal trend ($p = 0.07$). For Responsiveness, participants were less sensitive to latency changes. The only statistically significant difference was found between the 2 ms and 29 ms conditions ($p = 0.02$). The difference between 14.3 ms and 29 ms showed a marginal trend ($p = 0.07$), while the 2 ms condition was not rated significantly different from 23 ms ($p = 0.11$). While individual pairwise comparisons (Tukey-adjusted) lose power at these marginal levels, the main effect of latency remains statistically significant for both metrics when latency is modeled as a continuous or ordered categorical variable.

*Rankings.* In contrast to the ratings, the forced-choice ranking data (analyzed via Plackett-Luce) revealed a highly significant separation between the faster and slower latencies. Both 2 ms and 14.3 ms were significantly preferred over both 23 ms and 29 ms ($p < 0.001$ for all four cross-comparisons). There were no significant differences in ranking within the top pair (2 vs 14.3 ms, $p = 0.99$) or the bottom pair (23 vs 29 ms, $p = 0.86$). This indicates that while users struggle to quantify the difference on a rating scale, they reliably perceive the degradation in direct comparison, placing the threshold for ideal subjective preference between 14.3 and 23 ms.

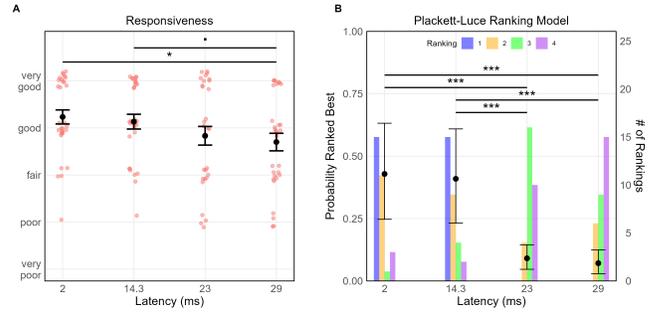

Fig. 9. **(A)** Responsiveness ratings provided after each block. The 2 ms condition was rated significantly higher than 29 ms ($p = 0.02$), while the 14.3 ms condition showed a marginal trend toward improvement over 29 ms ($p = 0.07$). Differences between 2 ms and 23 ms were not statistically significant ($p = 0.11$). All $p$-values reflect Tukey-adjusted pairwise comparisons. **(B)** Forced-choice rankings reveal a clear clustering effect: participants significantly preferred both 2 and 14.3 ms over the 23 and 29 ms conditions ($p < 0.001$), with no significant preference found within the lower latency (2 vs 14.3 ms) or higher latency (23 vs 29 ms) pairs.

## 5.3 Psychophysical Experiment

Having established the subjective preferences and performance trends in the ball catching task, we next sought to determine if these naturalistic results correlate with low-level perceptual sensitivity. To do this, we measured each participant's latency detection threshold using a standard two-interval forced-choice (2IFC) task. This allows us to compare individual psychophysical sensitivity directly against the acceptability and performance data collected in the previous section.

*5.3.1 Hardware.* Experiments were conducted using the zero-latency, head-tracked testbed architecture detailed by Guan et al. [2023]. The apparatus consists of a high-frequency (1000 Hz) rotary encoder that restricts head movement to a single axis (yaw), fixing the viewer's nominal eye positions relative to the center of rotation via a calibrated bite bar. Visual stimuli were displayed on an 88-inch OLED TV (OLED88ZXPUA) viewed from 57 cm, yielding a 125° × 94° field of view. A photodiode on the screen synchronized the display to a pair of shutter glasses for temporally-interlaced stereo presentation at 120 Hz. As established in prior characterization of this system, the baseline motion-to-photon latency is 26 ms. The renderer employs a velocity and acceleration-based forward prediction model to account for this 26 ms delay to achieve an effective end-to-end (e2e) latency of 0 ms. The effectiveness of this system was validated using both physical measurement with a photodiode to measure e2e latency from an encoder update (via button press) to display update, and expert perceptual evaluation by hand tuning the value for latency prediction to achieve the most apparent world-locked rendering quality during head motion. In both cases the measured e2e latency values were 26 ms.

*5.3.2 Psychophysical Task.* The scenes and stimuli used to measure latency detection thresholds are shown in Figure 10. We measured thresholds in an augmented reality (AR) scenario to validate the efficacy of each participant's bite bar calibration (as thresholds are expected to be significantly lower in AR) alongside thresholds in VR. In the AR scene, three physical posts are placed 10 cm in front of the display and three virtual spheres are rendered above them. Without latency the posts remain world-locked to the posts, but move relative to the posts if the user moves their head and the scene is rendered with added latency. In the VR scene we rendered content to fill a wide field of view, and the optical posts are removed so there are no real-world references to highlight the presence of latency. For both scenes, participants were instructed to fixate on a target while making a yaw head rotation, thereby inducing a vestibulo-ocular reflex (VOR) eye movement. Participants were not instructed to rotate at a specific velocity but were encouraged to rotate far and fast enough to sufficiently view the scene.

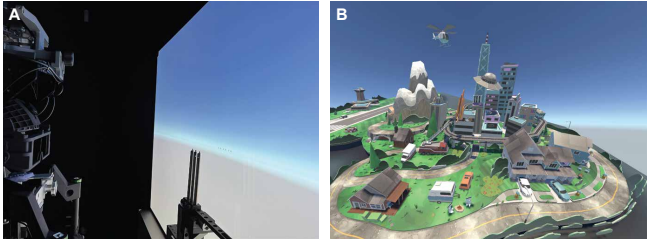

Fig. 10. Experimental scenes. Subjects performed experiments using a rotating chin rest while viewing virtual content on a TV display, which was positioned 57 cm from the CoR of their corneal apexes. **(A)** In the AR experiment, we presented three virtual spheres which hovered over three physical optical posts that served as world frames of reference. **(B)** In the VR experiment, we presented a 3D cartoon cityscape. This scene was more complex than the AR environment, with richer depth cue information such as occlusion, perspective, relative size, and volumetric shape.

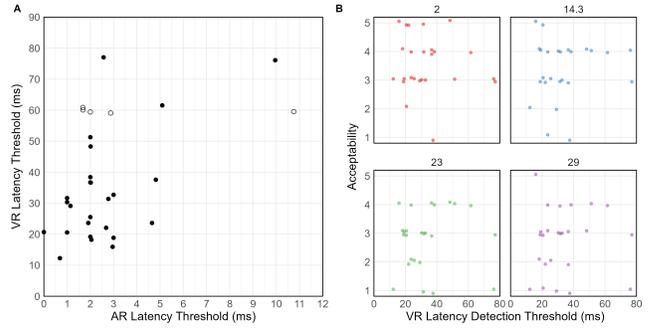

Fig. 11. **(A)** AR and VR latency detection thresholds for all 30 participants. AR thresholds were consistently low (2.8 ± 2.4 ms). In VR, five participants (open circles) exceeded the initial test range of 60 ms without reaching a threshold. VR detection thresholds for the remaining 25 participants is 33.6 ± 17.1 ms. **(B)** Scatter plots assessing the impact of VR thresholds on subjective Acceptability ratings. As described in Section 5.4, no significant relationship was found; participants with high sensitivity (low thresholds) did not rate the latencies differently than those with low sensitivity.

During the experiment, each trial consisted of two intervals: one without latency and another with a latency magnitude determined by an adaptive sampling algorithm (AEPsych [Owen et al. 2021]). Participants indicated via button press which of the two intervals appeared more stable. For all participants, the first five trials served as training, using large latency values and auditory feedback to familiarize them with the protocol. If a participant responded incorrectly during training, the experiment was restarted and instructions were reiterated.

Following the training phase, data collection trials were conducted without feedback. The initial batch of trials sampled the parameter space using Sobol sequences to ensure global coverage, while the remaining trials were adaptive, focusing on latencies near the subject's estimated threshold. For the AR scene, we collected 100 trials per participant (50 Sobol, 50 adaptive) within a range of 0 to 25 ms. For the VR scene, we collected 200 trials (50 Sobol, 150 adaptive) with latencies ranging from 0 to 60 ms (the upper bound was increased to 200 ms for some participants who struggled during the initial training trials).

*5.3.3 Data Analysis.* After the experiment, a logistic psychometric function was fit to the data using the Palamedes toolbox in MAT-LAB [Prins and Kingdom 2018]. The fit assumed a 50% guess rate and a 0.01% lapse rate. Threshold was defined as the latency from the psychometric fit where the subject correctly selected the 0-ms latency interval for 75% of the trials.

*5.3.4 Results.* In the AR scene, even minimal latency induces perceptible relative motion (spatial instability) between the real world and virtual spheres. Across all participants (n=30), the mean AR detection threshold was 2.8 ± 2.4 ms (Figure 11). To contextualize this sensitivity, we modeled the geometric error for an observer (61.4 mm IPD) performing sinusoidal head rotations (0.5 Hz, ±15.6°). We found that 1 ms of latency translates to a peak binocular disparity error of 29 arcseconds—a value consistent with known detection thresholds in relative motion tasks [Legge and Campbell 1981; Westheimer 1975].

As expected, participants were less sensitive to latency in the VR condition. Five participants in the initial cohort were unable to reliably discriminate intervals at the initial upper bound of 60 ms. Consequently, their true thresholds could not be determined. To accommodate this higher-than-expected variance, the stimulus range was extended to 200 ms for subsequent participants who demonstrated difficulty in successfully completing the training trials. For the subset of participants whose thresholds were successfully quantified (n=25), the mean VR detection threshold was 33.6 ± 17.1 ms.

Despite the difference in absolute magnitude between AR and VR thresholds, we found a significant positive correlation between the two modalities ($r(23) = 0.57, p = 0.003$). This suggests that while detection is more difficult in VR due to the lack of a real-world reference, individual perceptual sensitivity is consistent across conditions. Individual AR and VR latency detection thresholds are presented in Figure 11, with open circles indicating the five participants who threshold could not be measured within the initial 60 ms upper limit.

### 5.4 Correlating Thresholds to Ball Catching

We investigated whether individual differences in psychophysical detection thresholds could predict each participant's subjective evaluations and ranking accuracy in the ball-catching task, but did not find any meaningful correlations between them. We first examined the Pearson correlation between each participant's AR and VR detection thresholds and their ranking error, defined as the deviation of their subjective ranking from the true latency order (e.g., ranking 29 ms as "Best" corresponds to a maximum positional error of three). This analysis found no statistically significant relationship between either AR ($p = 0.71$) or VR ($p = 0.32$) detection thresholds and ranking performance.

We also included detection thresholds as a fixed effect in the cumulative link mixed models (CLMM) used to analyze Acceptability and Responsiveness Likert ratings. The model specification was:

$$\text{Rating} \sim \text{Latency} \times \text{Threshold} + (1 \mid \text{PID}) \qquad (5)$$

This interaction term tests whether the effect of latency on ratings is modulated by an individual's perceptual sensitivity (Figure 11B). We did not observe any significant interactions between detection thresholds and ratings, indicating that subjective ratings of responsiveness and acceptability were not meaningfully predicted by low-level latency detection sensitivity. We explore possible explanations in the Discussion.

## 6 DISCUSSION

*2 ms e2e Latency.* In both psychophysical and naturalistic evaluations, we find evidence that e2e latency targets for HMDs should be as low as 2 ms. However, the psychophysical results were derived from an AR scenario, where virtual content is compared against a physical reference. Initially, this target may seem to conflict with our video passthrough (Camsicle) studies. These are arguably more similar to VR because the entire visual field is subject to latency, rather than just a portion of it. Yet, unlike a purely VR scene, the participant's interaction with the physical world introduces potential sensory conflicts—such as proprioceptive mismatch during a ball catch—when the passthrough feed is delayed. Generally, the perceptual system is far more sensitive to relative differences than absolute errors [Mckee and Nakayama 1984; Stevenson et al. 1989]. This has implications for mixed reality systems that overlay virtual content onto a video passthrough stream which creates two distinct latency pathways: one for the video feed and another for the rendering pipeline. To minimize perceptual artifacts, it may be advantageous to synchronize these latencies—delaying the faster pipeline to match the slower one—rather than optimizing each path independently for the lowest possible latency.

*Detectability vs. Acceptability.* The relationship between the detectability and acceptability in user experience remains complex, and we did not find a reliable correlation between our participants' psychophysical thresholds (in AR or VR) and their subjective ratings or forced-choice rankings of e2e latency in Camsicle. While traditional psychophysics focuses on threshold detection—identifying the minimum perceptible latency—this metric does not always align with what users find acceptable in naturalistic scenarios. Recent methodological advances in psychophysics [Hong et al. 2025; Maloney and Yang 2003; Mikhailiuk et al. 2021; Owen et al. 2021] have expanded the scope of psychophysical characterization to include suprathreshold discrimination rather than simple detection. Characterizing this suprathreshold space may lead to more reliable links between psychophysical and naturalistic data. For example, an individual capable of distinguishing between 10 and 14 ms of latency may be more sensitive to quality differences in a naturalistic task than someone who can only discriminate between 10 and 20 ms. Capturing this 'sensitivity to change' rather than just 'sensitivity to presence' may yield stronger correlations with individual differences in more naturalistic settings.

*Subjective vs. Objective Metrics.* A notable finding is the divergence between subjective ratings and objective performance. Participants did not consistently rate or rank 2 ms latency as superior to 14.3 ms, despite performing better at lower latencies. This suggests that latency levels below the sensitivity threshold of survey-based measures (or potentially below perceptible thresholds) can still influence task performance. Recent work has also demonstrated that sub-threshold visual artifacts can also lead to increased visual discomfort [Levulis et al. 2025]. Therefore, relying solely on subjective feedback on quality may underestimate the impact of latency on user experience, underscoring the importance of incorporating measures of comfort and user performance in the evaluation of immersive systems.

*Study Limitations.* Several limitations should be considered when interpreting the results of this study. First, the headset used was tethered, which restricted participant movement and reduces ecological validity compared to untethered systems. Second, the headset's design is less ergonomic compared to modern, commercially available headsets due to its cantilevered weight distribution, potentially affecting user experience and performance. Third, the camera's 1 ms exposure time, while necessary for low-latency operation, led to noticeable judder when objects were not tracked by the user's eyes. These factors may have influenced both subjective ratings and objective performance, and future work could address these hardware constraints to improve comfort and realism.

## 7 CONCLUSION

We conducted a comprehensive psychophysical and naturalistic assessment of end-to-end (e2e) latency for head-mounted displays. We accomplished this by introducing a custom video passthrough headset capable of achieving latencies as low as 2 ms with a deterministic "render" and display pipeline. We used this headset to evaluate both subjective experience and objective performance across three distinct tasks and also measured psychophysical latency thresholds in a reference-grade test bed. We find both psychophysical thresholds and performance benefits for e2e latency in the 2 ms range. Notably, we found that subjective latency ratings plateaued at 14.3 ms, with no reported improvement at 2 ms. This divergence highlights the need for both objective and subjective approaches in study design to fully characterize the impacts of hardware and software on user experience.

# Perceptual Requirements for Low-Latency Head-Mounted Displays
Supplementary Material

ERIC PENNER, Reality Labs Research, Meta, USA
JOSEPHINE D'ANGELO, Reality Labs Research, Meta University of California, Berkeley, USA
CLINTON SMITH, Reality Labs Research, Meta, USA
NATHAN MATSUDA, Reality Labs Research, Meta, USA
NEETHAN SIVA, Reality Labs, Meta, USA
PHILLIP GUAN, Reality Labs Research, Meta, USA

## 1 CAMSICLE HARDWARE IMPLEMENTATION DETAILS

*Compute Hardware.* All our profiling and user studies were performed on a Lenovo P620 desktop, which contains an AMD ThreadRipper CPU with 32 cores, an Nvidia 3080 GPU, and the installed Alveo U45N network accelerator. Our compute pipeline is nonetheless very lightweight, and performs similar computations per pixel to stand-alone mobile SoCs. Additionally, since our camera-to-display distortion is fixed it is also amenable to implementation directly in a hardware display compositor.

*Display and Display Optics.* Our display optical pods consist of pancake lenses, utilizing 1.73" OLEDs with a resolution of 2768x3000. The total binocular FOV is 100° vertical and 110° horizontal at 30 PPD (pixels-per-degree). The displays are capable of 72hz and 90hz refresh rates with programmable low persistence rolling refresh.

*Camera Sensor and Lens.* For our camera sensor we used the AR2020 sensor from Onsemi, mounted on a low profile module with an interchangeable lens mount. These modules themselves are mounted on modular camera stalks, to enable simulation of arbitrary camera placement. The sensors utilize dual 4X mipi over micro-coax to reach the sensor's maximum spec of 5120x3840 10bit RAW, at 60hz. To run at 72hz, we crop the sensor vertically to 3160 rows, which very closely matches our display's PPD when using an appropriate fisheye lens. To run at higher rates binning must be used which halves the sensor resolution, and also results in higher latency due to mismatched scanning rates between camera and display.

*FPGA and Camera Driver.* Camera sensor pixels are written directly to host PC memory using two FPGAs and an optical fiber connection between them. A Xilinx XCKU5P FPGA on the headset expands 10 bit pixel data to 16 bits and places a microsecond timestamp in the last 32 bits of each row. Each row is streamed to an Alveo U45N accelerator in the PC, which contains an XCU26 Xilinx FPGA. On the PC, a custom camera driver locks memory for one full camera frame and looks up the raw memory address of each physical memory page. These raw memory addresses are provided to the FPGA, which is then able to performs scattered writes directly to the correct physical memory pages over the PCIE bus. Since a single camera frame is stored and reused, reads must synchronized to avoid tearing. Camera tearing can be detected by observing the timestamps of rows as they are read from memory (tearing presents as non-monotonic row times).

*Camera and Display Synchronization.* Synchronization between the display and camera hardware is achieved by wiring a vsync output pin on the display driver board into the FPGA, and using that as the camera frame trigger. A programmable phase offset is implemented to allow for a delay between camera trigger and display vsync. Triggering the camera earlier allows for buffering for distortion and scan timing differences.

*GPU Compositor.* Our GPU compositor operates in a tight loop which composites small slices just ahead of the display scanline. The display scan (and thus also the camera's scan - since they are synchronized) is inferred from the display's mode descriptor (or the EDID) and the time since the last vsync. To eliminate all copies and minimize CPU work in the compositor, we utilize persistent memory mapping (D3D11_MAP_WRITE_NO_OVERWRITE in Direct3D 11) which makes the system memory camera buffer directly readable from our GPU shaders while also writable from the FPGA. The result is that our compositor loop consists of a single constant buffer update and one or two draw calls per slice. Our compositing shader performs demosiacing similar to [McGuire 2009] followed by color and gamma correction. The exact slice size required is calculated on the fly, but we throttle the loop using a CPU busy-wait to keep dispatches to a minimum for the current latency target. When throttling the compositor to 0.1ms per slice, this corresponds to around 23 display rows per slice.

*Latency Jitter.* While our compositor is not very compute intensive, it is extremely latency sensitive when running at the lowest latency target. While an average compositing dispatch loop takes less than 0.02ms, a lead time in front of the display scan out must be added to handle the worst case possible dispatch time. Failing to render in time results in display tearing. Latency jitter can jump to 1-2ms on a non-real-time operating system, so to eliminate latency jitter we reserve a CPU core using Process Lasso's "reserved CPU sets" feature, in conjunction with using thread affinities and the highest thread priority. This is further improved by disabling hyper-threading, CPU C-states, and variable clock rates in both the BIOS and Windows power settings.

*High Speed Camera Latency Validation.* While our pipeline is capable of measuring per pixel passthrough latency using hardware row clocks, it relies on these clocks being accurate. While we implemented the camera pipeline ourselves, we don't have control of the operating system or GPU vsync clock. We validated our display and e2e latency using a 2000fps high speed camera. To create a reference light pulse, we trigger LEDs via parallel port pins (which have 4us round-trip latency), and validate that we can trigger display pixels all the way until 0.1ms in front of the beam (and conversely these draws "disappear" if issued 0.1ms behind the beam). See our supplementary video for examples of this high speed camera footage.

## 2 EXPERIMENT 1 ADDITIONAL DATA
### 2.1 Survey Questions

Subjective Questionnaire Items

| Question Item | Response Scale |
| --- | --- |
| **Responsiveness**<br>How would you rate the responsiveness of the passthrough to your head, body, or hand movements? | 1. Very Poor<br>2. Poor<br>3. Fair<br>4. Good<br>5. Very Good |
| **Acceptability**<br>If you had to use passthrough like this every day for tasks similar to the ones you just did, given the quality of the visual experience, how acceptable do you think it is? | 1. I would not use this<br>2. I might use this<br>3. I could use this if I needed to<br>4. I would use this<br>5. I would definitely use this |
| **Performance**<br>How successful were you in accomplishing what you were asked to do? | 1. Failure<br>2. –<br>3. –<br>4. –<br>5. Perfect |
| **Frustration**<br>How discouraged, irritated, and annoyed were you? | 1. Very Low<br>2. –<br>3. –<br>4. –<br>5. Very High |
| **Effort**<br>How hard did you work to accomplish your level of performance? | 1. Very Low<br>2. –<br>3. –<br>4. –<br>5. Very High |

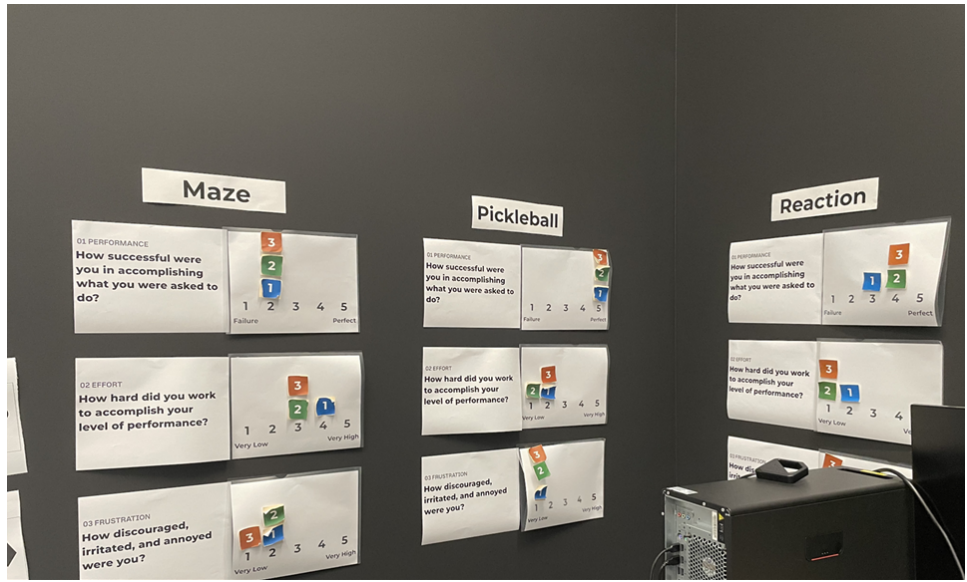

Fig. 1. Surveys for ratings of subjective performance, frustration, and effort. Participants were allowed to re-rate previous sessions based on their most recent experience.

## 2.2 Acceptability GLMM Analysis

Cumulative Link Mixed Model Results for Acceptability (Linear Trend)

| Predictor | Estimate | SE | z | p |
|---|---|---|---|---|
| *Fixed Effects* | | | | |
| Condition (Continuous) | -0.06 | 0.02 | -2.47 | .013 |
| *Random Effects* | | | | |
| | **Variance** | **SD** | | |
| PID (Intercept) | 5.60 | 2.37 | | |
| *Threshold Coefficients* | | | | |
| | **Estimate** | **SE** | | |
| 1\|2 | -4.25 | 0.87 | | |
| 2\|3 | -2.61 | 0.75 | | |
| 3\|4 | 0.58 | 0.64 | | |
| 4\|5 | 5.35 | 1.30 | | |

*Note.* $N_{PID} = 27$. Model fitted using Laplace approximation.

Cumulative Link Mixed Model Results for Acceptability (Ordered Categorical)

| Predictor | Estimate | SE | z | p |
|---|---|---|---|---|
| *Fixed Effects* | | | | |
| Linear Trend (L) | -1.11 | 0.44 | -2.51 | .012 |
| Quadratic Trend (Q) | -1.45 | 0.46 | -3.13 | .002 |
| *Random Effects* | | | | |
| | Variance | SD | | |
| PID (Intercept) | 7.72 | 2.78 | | |
| *Threshold Coefficients* | | | | |
| | Estimate | SE | | |
| 1\|2 | -3.88 | 0.84 | | |
| 2\|3 | -2.03 | 0.70 | | |
| 3\|4 | 1.64 | 0.68 | | |
| 4\|5 | 7.05 | 1.55 | | |

*Note.* $N_{PID} = 27$.

Pairwise Comparisons for Acceptability (Post-Hoc Contrasts)

| Contrast | Estimate | SE | z-ratio | $p_{adj}$ |
|---|---|---|---|---|
| Cond 2 – Cond 14.3 | -0.98 | 0.60 | -1.63 | .232 |
| Cond 2 – Cond 29 | 1.57 | 0.63 | 2.51 | .033 |
| Cond 14.3 – Cond 29 | 2.56 | 0.69 | 3.72 | < .001 |

*Note. p*-values are Tukey-adjusted for multiple comparisons.

## 2.3 Responsiveness GLMM Analysis

Cumulative Link Mixed Model Results for Responsiveness (Linear Trend)

| Predictor | Estimate | SE | z | p |
|---|---|---|---|---|
| *Fixed Effects* | | | | |
| Condition (Continuous) | -0.09 | 0.03 | -3.71 | < .001 |
| *Random Effects* | | | | |
| | Variance | SD | | |
| PID (Intercept) | 5.91 | 2.43 | | |
| *Threshold Coefficients* | | | | |
| | Estimate | SE | | |
| 1\|2 | -7.86 | 1.53 | | |
| 2\|3 | -6.30 | 1.20 | | |
| 3\|4 | -1.83 | 0.70 | | |
| 4\|5 | 1.33 | 0.68 | | |

*Note.* $N_{PID} = 27$. Model fitted using Laplace approximation.

Cumulative Link Mixed Model Results for Responsiveness (Ordered Categorical)

| Predictor | Estimate | SE | z | p |
|---|---|---|---|---|
| *Fixed Effects* | | | | |
| Linear Trend (L) | -1.89 | 0.48 | -3.96 | < .001 |
| Quadratic Trend (Q) | -1.05 | 0.45 | -2.32 | .020 |
| *Random Effects* | | | | |
| | Variance | SD | | |
| PID (Intercept) | 7.24 | 2.69 | | |
| *Threshold Coefficients* | | | | |
| | Estimate | SE | | |
| 1\|2 | -6.93 | 1.47 | | |
| 2\|3 | -5.29 | 1.05 | | |
| 3\|4 | -0.46 | 0.55 | | |
| 4\|5 | 2.97 | 0.79 | | |

*Note.* $N_{PID} = 27$.

Pairwise Comparisons for Responsiveness (Post-Hoc Contrasts)

| Contrast | Estimate | SE | z-ratio | $p_{adj}$ |
|---|---|---|---|---|
| Cond 2 – Cond 14.3 | 0.05 | 0.57 | 0.09 | .996 |
| Cond 2 – Cond 29 | 2.68 | 0.68 | 3.96 | < .001 |
| Cond 14.3 – Cond 29 | 2.63 | 0.72 | 3.65 | < .001 |

*Note.* *p*-values are Tukey-adjusted for multiple comparisons.

## 2.4 Plackett-Luce Ranking

One participant had to leave the study early so we only have ranking data from 26, rather than 27 participants.

Simultaneous Tests for General Linear Hypotheses (Plackett-Luce)

| Contrast | Estimate | SE | z-value | $p_{adj}$ |
|---|---|---|---|---|
| 2ms – 14.3ms | 0.24 | 0.37 | 0.64 | .799 |
| 2ms – 29ms | 1.16 | 0.40 | 2.88 | .011 |
| 14.3ms – 29ms | 0.93 | 0.40 | 2.32 | .053 |

*Note.* p-values are adjusted for multiple comparisons using the single-step method [Hothorn et al. 2008].

## 2.5 Objective Performance Measures

### 2.5.1 Ball Catching.

Generalized Linear Mixed Model Results for 'Caught' (Linear Trend)

| Predictor | Estimate | SE | z | p |
|---|---|---|---|---|
| *Fixed Effects* | | | | |
| (Intercept) | 2.40 | 0.28 | 8.54 | < .001 |
| Condition (Continuous) | -0.05 | 0.01 | -5.23 | < .001 |
| *Random Effects* | | | | |
| | Variance | SD | | |
| PID (Intercept) | 1.09 | 1.04 | | |

*Note.* $N_{obs} = 810, N_{PID} = 27$. Binomial family (logit link).

Generalized Linear Mixed Model Results for 'Caught' (Ordered Categorical)

| Predictor | Estimate | SE | z | p |
|---|---|---|---|---|
| *Fixed Effects* | | | | |
| (Intercept) | 1.71 | 0.23 | 7.37 | < .001 |
| Linear Trend (L) | -0.87 | 0.17 | -5.11 | < .001 |
| Quadratic Trend (Q) | -0.06 | 0.17 | -0.38 | .707 |
| *Random Effects* | | | | |
| | Variance | SD | | |
| PID (Intercept) | 1.09 | 1.04 | | |

*Note.* $N_{obs} = 810, N_{PID} = 27$.

Pairwise Comparisons for 'Caught' (Post-Hoc Contrasts)

| Contrast | Odds Ratio | SE | z-ratio | $p_{adj}$ |
|---|---|---|---|---|
| Cond 2 / Cond 14.3 | 1.71 | 0.43 | 2.14 | .081 |
| Cond 2 / Cond 29 | 3.42 | 0.82 | 5.11 | < .001 |
| Cond 14.3 / Cond 29 | 2.00 | 0.44 | 3.15 | .005 |

*Note.* *p*-values are Tukey-adjusted for multiple comparisons.

### 2.5.2 Timed Maze.

Generalized Linear Mixed Model Results for 'Total Errors' (Linear Trend)

| Predictor | Estimate | SE | z | p |
|---|---|---|---|---|
| *Fixed Effects* | | | | |
| (Intercept) | 2.19 | 0.27 | 7.99 | < .001 |
| Condition (Continuous) | 0.01 | 0.002 | 3.43 | < .001 |
| Time | -0.04 | 0.02 | -2.35 | .019 |
| *Random Effects* | | | | |
| | Variance | SD | | |
| PID (Intercept) | 0.13 | 0.37 | | |

*Note.* $N_{obs} = 243, N_{PID} = 27$. Poisson family (log link).

8 • Eric Penner, Josephine D'Angelo, Clinton Smith, Nathan Matsuda, Neethan Siva, and Phillip GuanGeneralized Linear Mixed Model Results for 'Total Errors' (Ordered Categorical)

| Predictor | Estimate | SE | z | p |
|---|---|---|---|---|
| *Fixed Effects* | | | | |
| (Intercept) | 2.31 | 0.27 | 8.52 | < .001 |
| Linear Trend (L) | 0.15 | 0.05 | 3.39 | < .001 |
| Quadratic Trend (Q) | 0.02 | 0.05 | 0.41 | .685 |
| Time | -0.04 | 0.02 | -2.35 | .019 |
| *Random Effects* | | | | |
| | Variance | SD | | |
| PID (Intercept) | 0.13 | 0.37 | | |

*Note.* $N_{obs} = 243, N_{PID} = 27$.

Pairwise Comparisons for 'Total Errors' (Post-Hoc Contrasts)

| Contrast | Rate Ratio | SE | z-ratio | $p_{adj}$ |
|---|---|---|---|---|
| Cond 2 / Cond 14.3 | 0.92 | 0.06 | -1.31 | .392 |
| Cond 2 / Cond 29 | 0.80 | 0.05 | -3.39 | .002 |
| Cond 14.3 / Cond 29 | 0.88 | 0.06 | -2.10 | .090 |

*Note. p*-values are Tukey-adjusted for multiple comparisons.

### 2.5.3 Reaction Time.

Generalized Linear Mixed Model Results for Reaction Time (Linear Trend)

| Predictor | Estimate | SE | t | p |
|---|---|---|---|---|
| *Fixed Effects* | | | | |
| (Intercept) | -1.44 | 0.03 | -45.45 | < .001 |
| Condition (Scaled) | 4.77 | 0.66 | 7.22 | < .001 |
| *Random Effects* | | | | |
| | Variance | SD | | |
| PID (Intercept) | 0.006 | 0.08 | | |
| Residual | 0.026 | 0.16 | | |

*Note.* $N_{obs} = 405, N_{PID} = 27$. Gamma family (log link).

Generalized Linear Mixed Model Results for Reaction Time (Ordered Categorical)

| Predictor | Estimate | SE | t | p |
|---|---|---|---|---|
| *Fixed Effects* | | | | |
| (Intercept) | -1.37 | 0.03 | -45.49 | < .001 |
| Linear Trend (L) | 0.09 | 0.01 | 7.20 | < .001 |
| Quadratic Trend (Q) | 0.01 | 0.01 | 0.55 | .581 |
| *Random Effects* | | | | |
| | Variance | SD | | |
| PID (Intercept) | 0.006 | 0.08 | | |
| Residual | 0.026 | 0.16 | | |

*Note.* $N_{obs} = 405$, $N_{PID} = 27$.

Pairwise Comparisons for Reaction Time (Post-Hoc Contrasts)

| Contrast | Ratio | SE | z-ratio | $p_{adj}$ |
|---|---|---|---|---|
| Cond 2 / Cond 14.3 | 0.95 | 0.02 | -3.12 | .005 |
| Cond 2 / Cond 29 | 0.88 | 0.02 | -7.20 | < .001 |
| Cond 14.3 / Cond 29 | 0.93 | 0.02 | -4.08 | < .001 |

*Note. p*-values are Tukey-adjusted for multiple comparisons.

## 2.6 Subjective Performance, Frustration, and Effort

### 2.6.1 Subjective Performance.

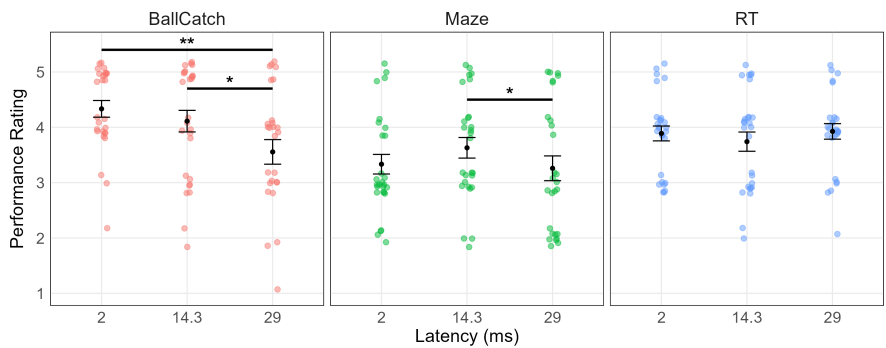

Fig. 2. Participant ratings of subjective performance across tasks.

Pairwise Comparisons for Performance (All Tasks)

| Contrast | Estimate | SE | z-ratio | $p_{adj}$ |
|---|---|---|---|---|
| **Ball Catch Task** | | | | |
| Cond 2 – Cond 14.3 | 0.61 | 0.59 | 1.02 | .563 |
| Cond 2 – Cond 29 | 2.25 | 0.64 | 3.49 | .001 |
| Cond 14.3 – Cond 29 | 1.64 | 0.60 | 2.71 | .018 |
| **Maze Task** | | | | |
| Cond 2 – Cond 14.3 | -1.29 | 0.65 | -1.99 | .114 |
| Cond 2 – Cond 29 | 0.50 | 0.64 | 0.78 | .717 |
| Cond 14.3 – Cond 29 | 1.79 | 0.69 | 2.60 | .025 |
| **Reaction Time Task** | | | | |
| Cond 2 – Cond 14.3 | 0.81 | 0.66 | 1.22 | .442 |
| Cond 2 – Cond 29 | -0.20 | 0.65 | -0.31 | .950 |
| Cond 14.3 – Cond 29 | -1.01 | 0.67 | -1.50 | .290 |

*Note. p*-values are Tukey-adjusted.

### 2.6.2 Frustration.

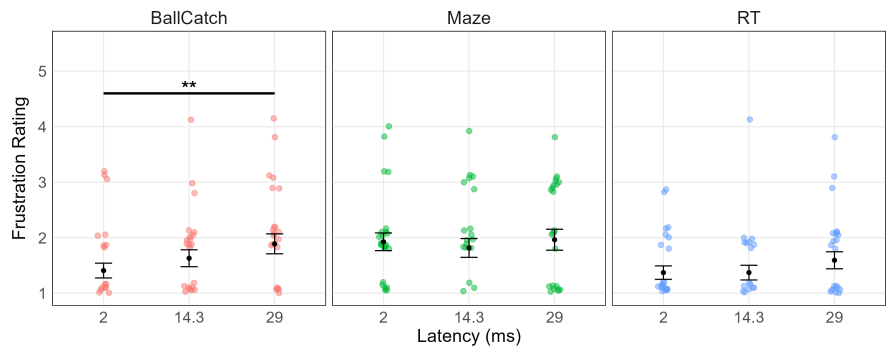

Fig. 3. Participant ratings of subjective frustration across tasks.

Pairwise Comparisons for Frustration (All Tasks)

| Contrast | Estimate | SE | z-ratio | $p_{adj}$ |
|---|---|---|---|---|
| **Ball Catch Task** | | | | |
| Cond 2 – Cond 14.3 | -1.20 | 0.72 | -1.66 | .220 |
| Cond 2 – Cond 29 | -2.37 | 0.77 | -3.07 | .006 |
| Cond 14.3 – Cond 29 | -1.17 | 0.66 | -1.78 | .177 |
| **Maze Task** | | | | |
| Cond 2 – Cond 14.3 | 0.49 | 0.60 | 0.82 | .693 |
| Cond 2 – Cond 29 | -0.00 | 0.60 | -0.01 | 1.000 |
| Cond 14.3 – Cond 29 | -0.49 | 0.61 | -0.80 | .703 |
| **Reaction Time Task** | | | | |
| Cond 2 – Cond 14.3 | 0.13 | 0.82 | 0.16 | .986 |
| Cond 2 – Cond 29 | -1.44 | 0.78 | -1.86 | .151 |
| Cond 14.3 – Cond 29 | -1.57 | 0.82 | -1.92 | .134 |

*Note. p*-values are Tukey-adjusted.

### 2.6.3 Effort.

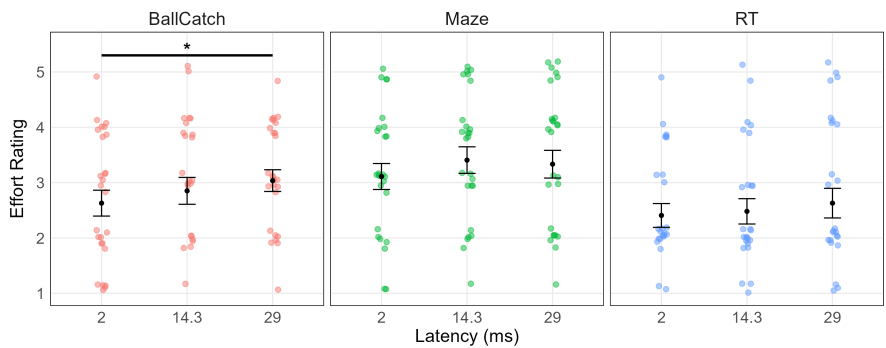

Fig. 4. Participant ratings of subjective effort across tasks.

Pairwise Comparisons for Effort (All Tasks)

| Contrast | Estimate | SE | z-ratio | $p_{adj}$ |
|---|---|---|---|---|
| **Ball Catch Task** | | | | |
| Cond 2 – Cond 14.3 | -0.75 | 0.58 | -1.28 | .407 |
| Cond 2 – Cond 29 | -1.42 | 0.59 | -2.40 | .044 |
| Cond 14.3 – Cond 29 | -0.68 | 0.56 | -1.20 | .452 |
| **Maze Task** | | | | |
| Cond 2 – Cond 14.3 | -1.25 | 0.63 | -1.96 | .122 |
| Cond 2 – Cond 29 | -0.95 | 0.62 | -1.53 | .275 |
| Cond 14.3 – Cond 29 | 0.29 | 0.61 | 0.48 | .881 |
| **Reaction Time Task** | | | | |
| Cond 2 – Cond 14.3 | -0.27 | 0.76 | -0.36 | .932 |
| Cond 2 – Cond 29 | -1.29 | 0.80 | -1.62 | .239 |
| Cond 14.3 – Cond 29 | -1.02 | 0.78 | -1.30 | .394 |

*Note. p*-values are Tukey-adjusted.

## 2.7 Additional Plots

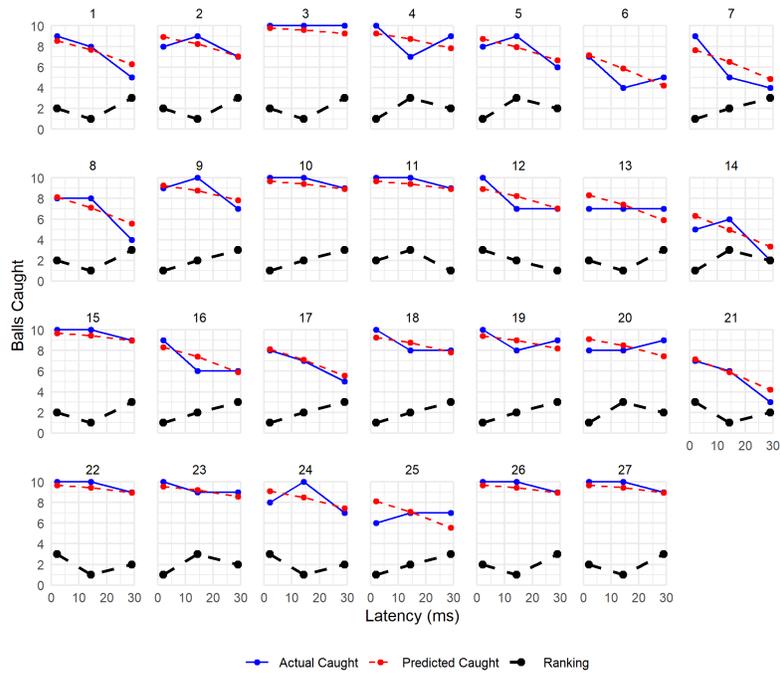

Fig. 5. Model fit for catch performance across latency for each individual.

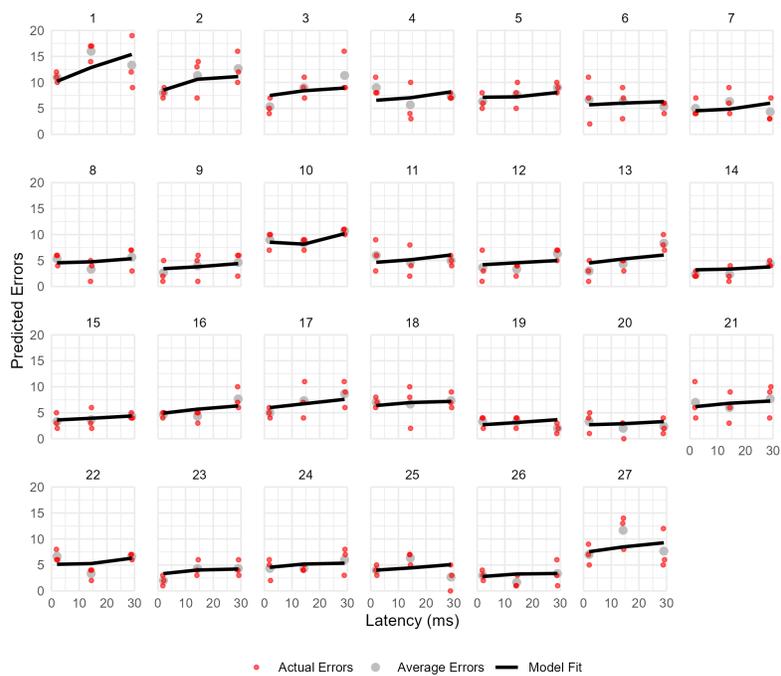

Fig. 6. Model fit for catch performance across latency for each individual.

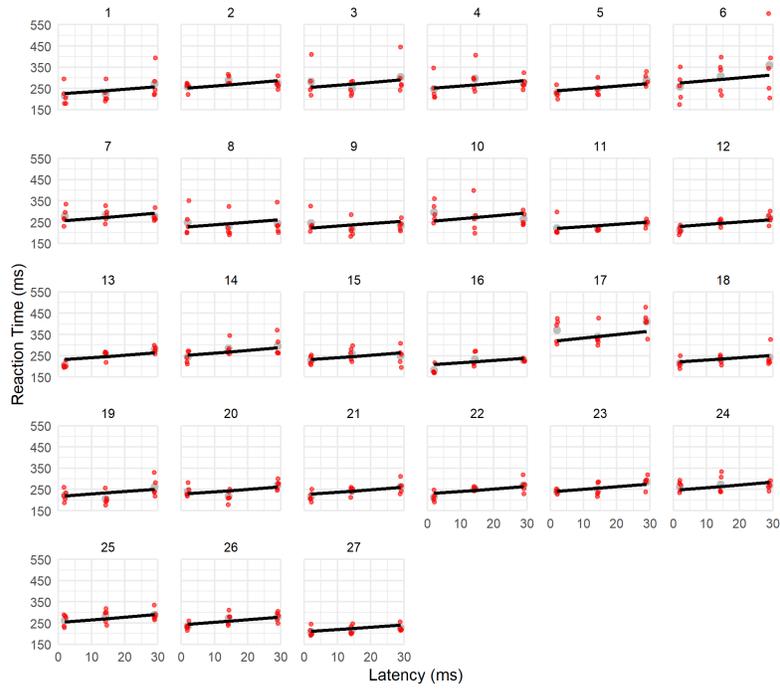

Fig. 7. Fixed and random effects per individual.

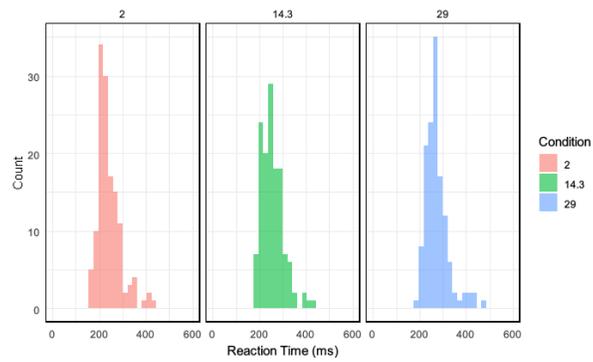

Fig. 8. Right skew RT distributions for Gamma error distribution.

## 3 EXPERIMENT 2 ADDITIONAL DATA
### 3.1 Psychometric Functions

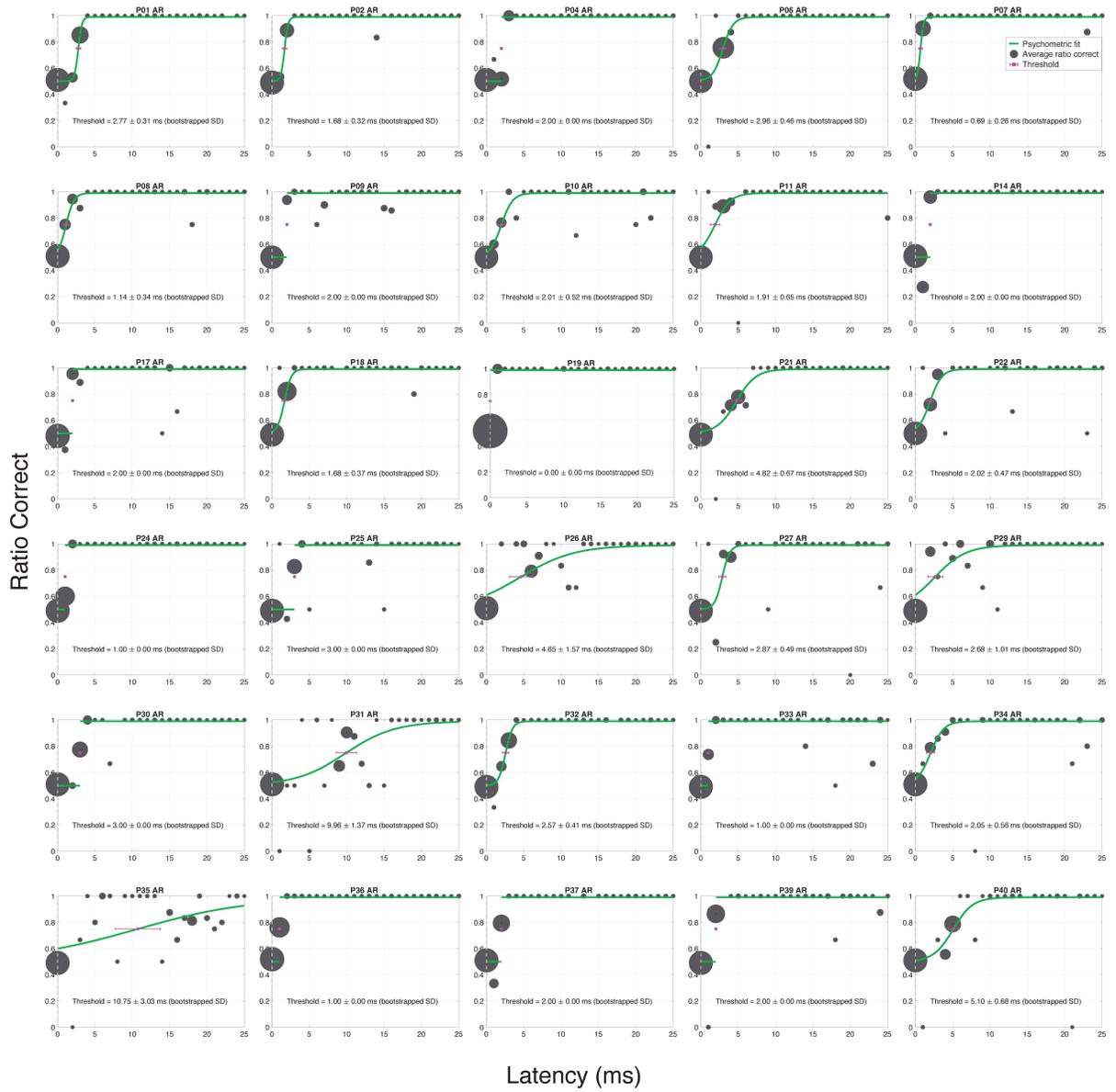

Fig. 9. Psychometric functions for all 30 participants for AR data.

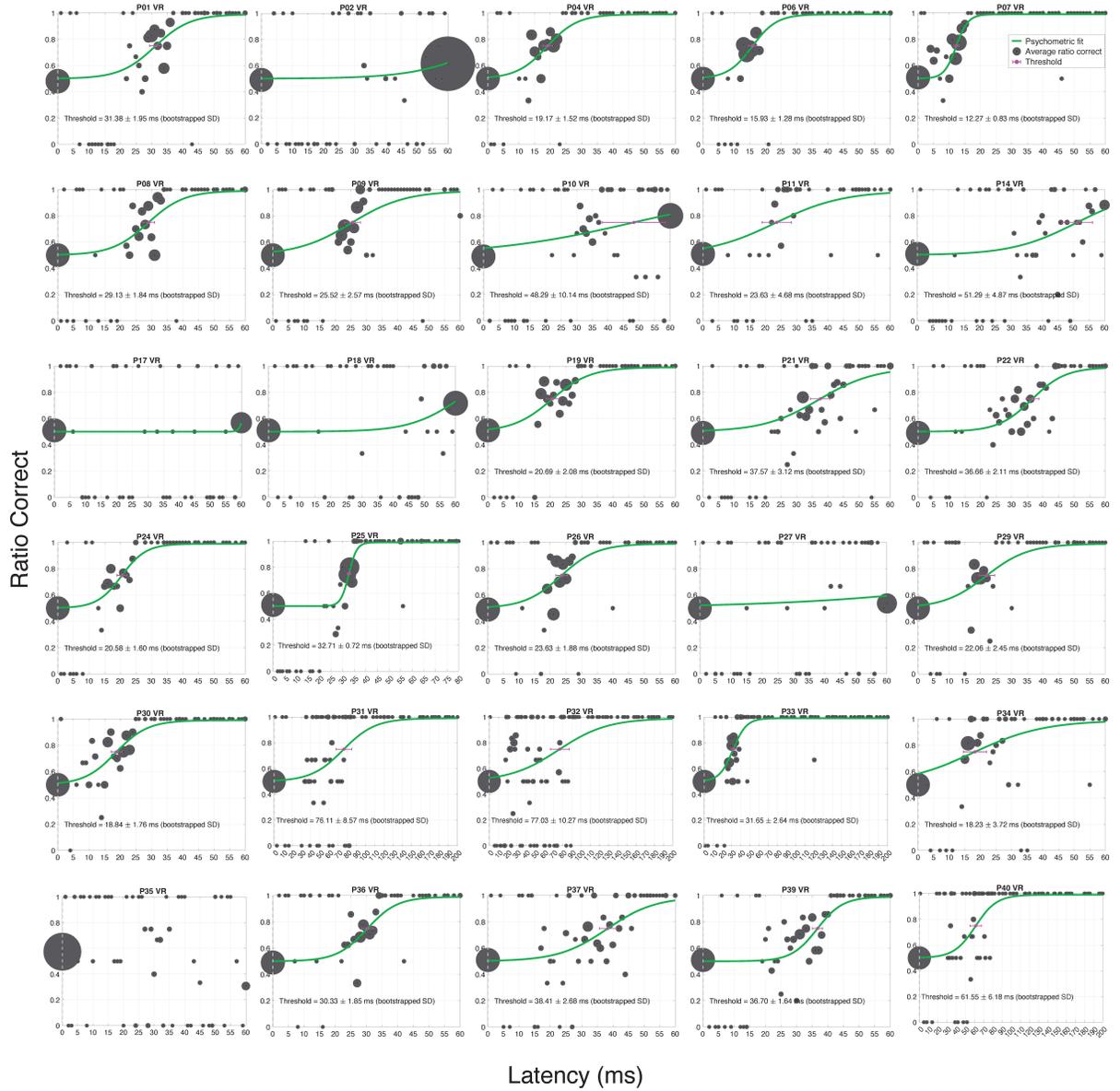

Fig. 10. Psychometric fits to all 30 participants for VR data.

Perceptual Requirements for Low-Latency Head-Mounted Displays • 17## 3.2 Acceptability GLMM Analysis

Table 1. Cumulative Link Mixed Model Results (Linear Trend)

| Predictor | Estimate | SE | z | p |
|---|---|---|---|---|
| *Fixed Effects* | | | | |
| Condition (Continuous) | -0.08 | 0.02 | -3.96 | < .001 |
| *Random Effects* | | | | |
| | Variance | SD | | |
| PID (Intercept) | 4.33 | 2.08 | | |
| *Threshold Coefficients* | | | | |
| | Estimate | SE | | |
| 1\|2 | -4.84 | 0.76 | | |
| 2\|3 | -3.27 | 0.66 | | |
| 3\|4 | -0.15 | 0.56 | | |
| 4\|5 | 2.95 | 0.68 | | |

*Note.* $N_{PID} = 30$.

Table 2. Cumulative Link Mixed Model Results (Ordered Categorical)

| Predictor | Estimate | SE | z | p |
|---|---|---|---|---|
| *Fixed Effects* | | | | |
| Linear Trend (L) | -1.65 | 0.41 | -4.02 | < .001 |
| Quadratic Trend (Q) | 0.00 | 0.37 | 0.01 | .992 |
| Cubic Trend (C) | 0.41 | 0.36 | 1.13 | .258 |
| *Random Effects* | | | | |
| | Variance | SD | | |
| PID (Intercept) | 4.44 | 2.11 | | |
| *Threshold Coefficients* | | | | |
| | Estimate | SE | | |
| 1\|2 | -3.52 | 0.60 | | |
| 2\|3 | -1.94 | 0.50 | | |
| 3\|4 | 1.23 | 0.48 | | |
| 4\|5 | 4.37 | 0.70 | | |

*Note.* $N_{PID} = 30$.

Table 3. Pairwise Comparisons (Post-Hoc Contrasts)

| Contrast | Estimate | SE | z-ratio | $p_{adj}$ |
|---|---|---|---|---|
| Cond 2 – Cond 14.3 | 0.37 | 0.53 | 0.70 | .896 |
| Cond 2 – Cond 23 | 1.66 | 0.55 | 3.04 | .013 |
| Cond 2 – Cond 29 | 2.02 | 0.57 | 3.56 | .002 |
| Cond 14.3 – Cond 23 | 1.29 | 0.53 | 2.45 | .068 |
| Cond 14.3 – Cond 29 | 1.65 | 0.54 | 3.04 | .013 |
| Cond 23 – Cond 29 | 0.36 | 0.51 | 0.72 | .890 |

*Note. p*-values are Tukey-adjusted for multiple comparisons.

## 3.3 Responsiveness GLMM Analysis

Table 4. Cumulative Link Mixed Model Results for Responsiveness (Linear Trend)

| Predictor | Estimate | SE | $z$ | $p$ |
|---|---|---|---|---|
| *Fixed Effects* | | | | |
| Condition (Continuous) | -0.06 | 0.02 | -3.16 | .002 |
| *Random Effects* | | | | |
| | Variance | SD | | |
| PID (Intercept) | 5.08 | 2.25 | | |
| *Threshold Coefficients* | | | | |
| | Estimate | SE | | |
| 2\|3 | -5.04 | 0.81 | | |
| 3\|4 | -2.75 | 0.67 | | |
| 4\|5 | -0.26 | 0.59 | | |

*Note.* $N_{PID} = 30$.

Table 5. Cumulative Link Mixed Model Results for Responsiveness (Ordered Categorical)

| Predictor | Estimate | SE | $z$ | $p$ |
|---|---|---|---|---|
| *Fixed Effects* | | | | |
| Linear Trend (L) | -1.33 | 0.41 | -3.24 | .001 |
| Quadratic Trend (Q) | -0.04 | 0.39 | -0.09 | .926 |
| Cubic Trend (C) | 0.30 | 0.39 | 0.76 | .448 |
| *Random Effects* | | | | |
| | Variance | SD | | |
| PID (Intercept) | 5.21 | 2.28 | | |
| *Threshold Coefficients* | | | | |
| | Estimate | SE | | |
| 2\|3 | -3.99 | 0.67 | | |
| 3\|4 | -1.67 | 0.54 | | |
| 4\|5 | 0.85 | 0.50 | | |

*Note.* $N_{PID} = 30$.

Table 6. Pairwise Comparisons for Responsiveness (Post-Hoc Contrasts)

| Contrast | Estimate | SE | $z$-ratio | $p_{adj}$ |
|---|---|---|---|---|
| Cond 2 – Cond 14.3 | 0.29 | 0.56 | 0.52 | .954 |
| Cond 2 – Cond 23 | 1.29 | 0.57 | 2.24 | .112 |
| Cond 2 – Cond 29 | 1.65 | 0.57 | 2.90 | .020 |
| Cond 14.3 – Cond 23 | 0.99 | 0.56 | 1.76 | .293 |
| Cond 14.3 – Cond 29 | 1.36 | 0.56 | 2.42 | .074 |
| Cond 23 – Cond 29 | 0.37 | 0.54 | 0.68 | .903 |

*Note.* $p$-values are Tukey-adjusted for multiple comparisons.

## 3.4 Plackett-Luce Ranking

Simultaneous Tests for General Linear Hypotheses (Plackett-Luce Model)

| Contrast | Estimate | SE | z-value | $Pr(>|z|)$ |
|---|---|---|---|---|
| Cond 2 – Cond 14.3 | 0.04 | 0.35 | 0.11 | .999 |
| Cond 2 – Cond 23 | 1.50 | 0.37 | 4.02 | < .001 |
| Cond 2 – Cond 29 | 1.76 | 0.39 | 4.47 | < .001 |
| Cond 14.3 – Cond 23 | 1.46 | 0.37 | 3.97 | < .001 |
| Cond 14.3 – Cond 29 | 1.72 | 0.39 | 4.40 | < .001 |
| Cond 23 – Cond 29 | 0.26 | 0.34 | 0.78 | .865 |

*Note.* p-values are adjusted for multiple comparisons using the single-step method [Hothorn et al. 2008].

## 3.5 Objective Ball-Catching Performance

Statistical analysis for GLMMs using continuous, ordered categorical, and post-hoc pairwise comparisons.

Table 7. Generalized Linear Mixed Model Results for 'Caught' (Linear Trend)

| Predictor | Estimate | SE | z | p |
|---|---|---|---|---|
| *Fixed Effects* | | | | |
| (Intercept) | 2.61 | 0.28 | 9.22 | < .001 |
| Condition (Continuous) | -0.08 | 0.01 | -9.33 | < .001 |
| *Random Effects* | | | | |
| | **Variance** | **SD** | | |
| PID (Intercept) | 1.34 | 1.16 | | |

*Note.* $N_{obs} = 1200$, $N_{PID} = 30$. Binomial family (logit link).

Table 8. Generalized Linear Mixed Model Results for 'Caught' (Ordered Categorical)

| Predictor | Estimate | SE | z | p |
|---|---|---|---|---|
| *Fixed Effects* | | | | |
| (Intercept) | 1.31 | 0.23 | 5.70 | < .001 |
| Linear Trend (L) | -1.58 | 0.17 | -9.11 | < .001 |
| Quadratic Trend (Q) | 0.55 | 0.16 | 3.51 | < .001 |
| Cubic Trend (C) | 0.01 | 0.14 | 0.10 | .919 |
| *Random Effects* | | | | |
| | **Variance** | **SD** | | |
| PID (Intercept) | 1.35 | 1.16 | | |

*Note.* $N_{obs} = 1200$, $N_{PID} = 30$.

Table 9. Pairwise Comparisons for 'Caught' (Post-Hoc Contrasts)

| Contrast | Odds Ratio | SE | z-ratio | $p_{adj}$ |
|---|---|---|---|---|
| Cond 2 / Cond 14.3 | 3.47 | 0.88 | 4.92 | < .001 |
| Cond 2 / Cond 23 | 7.16 | 1.78 | 7.94 | < .001 |
| Cond 2 / Cond 29 | 8.25 | 2.05 | 8.51 | < .001 |
| Cond 14.3 / Cond 23 | 2.07 | 0.41 | 3.64 | .002 |
| Cond 14.3 / Cond 29 | 2.38 | 0.47 | 4.36 | < .001 |
| Cond 23 / Cond 29 | 1.15 | 0.22 | 0.76 | .873 |

*Note. p*-values are Tukey-adjusted for multiple comparisons.

## 3.6 Subjective Performance, Frustration, and Effort

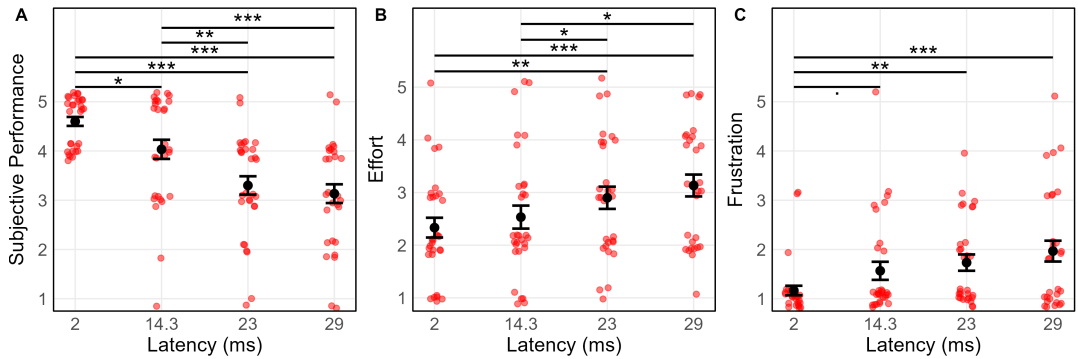

Fig. 11. Likert ratings shown with standard error for subjective ratings of performance, effort, and frustration. Note: p-values are adjusted for multiple comparison.

### 3.6.1 Subjective Performance.

Table 10. Cumulative Link Mixed Model Results for Performance (Linear Trend)

| Predictor | Estimate | SE | z | p |
|---|---|---|---|---|
| *Fixed Effects* | | | | |
| Condition (Continuous) | -0.15 | 0.02 | -6.21 | < .001 |
| *Random Effects* | | | | |
| | **Variance** | **SD** | | |
| PID (Intercept) | 1.63 | 1.28 | | |
| *Threshold Coefficients* | | | | |
| | **Estimate** | **SE** | | |
| 1\|2 | -7.12 | 0.90 | | |
| 2\|3 | -5.58 | 0.76 | | |
| 3\|4 | -3.70 | 0.62 | | |
| 4\|5 | -1.11 | 0.46 | | |

*Note.* $N_{PID}$ = 30.

Table 11. Cumulative Link Mixed Model Results for Performance (Ordered Categorical)

| Predictor | Estimate | SE | z | p |
|---|---|---|---|---|
| *Fixed Effects* | | | | |
| Linear Trend (L) | -3.06 | 0.49 | -6.20 | < .001 |
| Quadratic Trend (Q) | 0.63 | 0.38 | 1.66 | .096 |
| Cubic Trend (C) | 0.40 | 0.37 | 1.08 | .279 |
| *Random Effects* | | | | |
| | Variance | SD | | |
| PID (Intercept) | 1.69 | 1.30 | | |
| *Threshold Coefficients* | | | | |
| | Estimate | SE | | |
| 1\|2 | -4.57 | 0.64 | | |
| 2\|3 | -3.02 | 0.47 | | |
| 3\|4 | -1.13 | 0.36 | | |
| 4\|5 | 1.50 | 0.38 | | |

Note. $N_{PID}$ = 30.

Table 12. Pairwise Comparisons for Performance (Post-Hoc Contrasts)

| Contrast | Estimate | SE | z-ratio | $p_{adj}$ |
|---|---|---|---|---|
| Cond 2 – Cond 14.3 | 1.64 | 0.59 | 2.81 | .026 |
| Cond 2 – Cond 23 | 3.54 | 0.65 | 5.47 | < .001 |
| Cond 2 – Cond 29 | 3.94 | 0.68 | 5.83 | < .001 |
| Cond 14.3 – Cond 23 | 1.90 | 0.55 | 3.47 | .003 |
| Cond 14.3 – Cond 29 | 2.29 | 0.57 | 4.04 | < .001 |
| Cond 23 – Cond 29 | 0.39 | 0.49 | 0.81 | .852 |

Note. *p*-values are Tukey-adjusted for multiple comparisons.

### 3.6.2 Effort.

Table 13. Cumulative Link Mixed Model Results for Effort (Linear Trend)

| Predictor | Estimate | SE | z | p |
|---|---|---|---|---|
| *Fixed Effects* | | | | |
| Condition (Continuous) | 0.10 | 0.02 | 4.55 | < .001 |
| *Random Effects* | | | | |
| | Variance | SD | | |
| PID (Intercept) | 9.14 | 3.02 | | |
| *Threshold Coefficients* | | | | |
| | Estimate | SE | | |
| 1\|2 | -2.46 | 0.76 | | |
| 2\|3 | 1.36 | 0.71 | | |
| 3\|4 | 3.80 | 0.82 | | |
| 4\|5 | 6.72 | 1.08 | | |

Note. $N_{PID}$ = 30.

Table 14. Cumulative Link Mixed Model Results for Effort (Ordered Categorical)

| Predictor | Estimate | SE | z | p |
|---|---|---|---|---|
| *Fixed Effects* | | | | |
| Linear Trend (L) | 2.05 | 0.44 | 4.65 | < .001 |
| Quadratic Trend (Q) | 0.05 | 0.38 | 0.13 | .893 |
| Cubic Trend (C) | -0.36 | 0.38 | -0.97 | .334 |
| *Random Effects* | | | | |
|  | Variance | SD | | |
| PID (Intercept) | 9.51 | 3.08 | | |
| *Threshold Coefficients* | | | | |
|  | Estimate | SE | | |
| 1\|2 | -4.21 | 0.79 | | |
| 2\|3 | -0.33 | 0.64 | | |
| 3\|4 | 2.17 | 0.68 | | |
| 4\|5 | 5.14 | 0.92 | | |

*Note.* $N_{PID}$ = 30.

Table 15. Pairwise Comparisons for Effort (Post-Hoc Contrasts)

| Contrast | Estimate | SE | z-ratio | $p_{adj}$ |
|---|---|---|---|---|
| Cond 2 – Cond 14.3 | -0.54 | 0.55 | -0.99 | .756 |
| Cond 2 – Cond 23 | -1.95 | 0.58 | -3.37 | .004 |
| Cond 2 – Cond 29 | -2.59 | 0.61 | -4.27 | < .001 |
| Cond 14.3 – Cond 23 | -1.41 | 0.55 | -2.54 | .054 |
| Cond 14.3 – Cond 29 | -2.05 | 0.58 | -3.53 | .002 |
| Cond 23 – Cond 29 | -0.64 | 0.53 | -1.22 | .616 |

*Note. p*-values are Tukey-adjusted for multiple comparisons.

*3.6.3 Frustration.*

Table 16. Cumulative Link Mixed Model Results for Frustration (Linear Trend)

| Predictor | Estimate | SE | z | p |
|---|---|---|---|---|
| *Fixed Effects* | | | | |
| Condition (Continuous) | 0.13 | 0.03 | 4.30 | < .001 |
| *Random Effects* | | | | |
|  | Variance | SD | | |
| PID (Intercept) | 4.49 | 2.12 | | |
| *Threshold Coefficients* | | | | |
|  | Estimate | SE | | |
| 1\|2 | 3.45 | 0.83 | | |
| 2\|3 | 4.94 | 0.95 | | |
| 3\|4 | 7.41 | 1.23 | | |
| 4\|5 | 8.87 | 1.44 | | |

*Note.* $N_{PID}$ = 30.

Table 17. Cumulative Link Mixed Model Results for Frustration (Ordered Categorical)

| Predictor | Estimate | SE | z | p |
|---|---|---|---|---|
| *Fixed Effects* | | | | |
| Linear Trend (L) | 2.72 | 0.65 | 4.20 | < .001 |
| Quadratic Trend (Q) | -0.81 | 0.52 | -1.57 | .116 |
| Cubic Trend (C) | 0.24 | 0.45 | 0.54 | .587 |
| *Random Effects* | | | | |
| | **Variance** | **SD** | | |
| PID (Intercept) | 4.56 | 2.14 | | |
| *Threshold Coefficients* | | | | |
| | **Estimate** | **SE** | | |
| 1\|2 | 1.26 | 0.53 | | |
| 2\|3 | 2.77 | 0.62 | | |
| 3\|4 | 5.25 | 0.91 | | |
| 4\|5 | 6.70 | 1.14 | | |

Note. $N_{PID}$ = 30.

Table 18. Pairwise Comparisons for Frustration (Post-Hoc Contrasts)

| Contrast | Estimate | SE | z-ratio | $p_{adj}$ |
|---|---|---|---|---|
| Cond 2 – Cond 14.3 | -2.24 | 0.89 | -2.53 | .056 |
| Cond 2 – Cond 23 | -3.13 | 0.89 | -3.52 | .002 |
| Cond 2 – Cond 29 | -3.76 | 0.93 | -4.06 | < .001 |
| Cond 14.3 – Cond 23 | -0.89 | 0.62 | -1.43 | .478 |
| Cond 14.3 – Cond 29 | -1.51 | 0.64 | -2.36 | .085 |
| Cond 23 – Cond 29 | -0.62 | 0.56 | -1.11 | .685 |

Note. *p*-values are Tukey-adjusted for multiple comparisons.

## 4 SIMULATING THE LATENCY TESTBED

We simulated the WLR-capable apparatus using the following parameters:
- Head rotated: −25° to +25°.
- Front of eye: 9.12 cm from the head center of rotation (CoR).
- Eye CoR: 1.2 cm behind the front of the eye.
- Entrance pupil position (where the camera is placed): 1.0 cm in front of the eye CoR.
- TV display: 57 cm from the front of the eye.
- Spheres: 20 cm from the front of the eye.
- Interpupillary distance: 60 mm.
- Camera separation: 60 mm.

To verify the simulation, we measured 1) the magnitude of disparity when the head faces forward (0°), 2) the total translation on the screen for a monocular image with 0-ms latency during head rotation, and 3) the total translation on the screen for a monocular image with 200-ms latency during head rotation. We compared the simulation's values to those measured on the physical TV display. Note that the TV display measurements were estimated using a metric ruler and are therefore noisier than the simulation predictions.

We tested two conditions: 1) where the ray traces were drawn from the eye CoR and 2) where the ray traces were drawn from the visual axis of the entrance pupil positions, after accounting for rotation due to vestibular ocular reflex (VOR) and angle kappa (5°). Table 19 shows the simulation predictions compared to the actual TV measurements for each condition. The simulation predictions closely match the actual measurements on the TV display.

|  | CoR | | Visual Axis | |
| --- | --- | --- | --- | --- |
|  | Sim (cm) | TV (cm) | Sim (cm) | TV (cm) |
| **Head forward** | 10.47 | 10.5 | 10.75 | 10.75 |
| **0-ms latency** | 10.79 | 10.7 | 10.69 | 10.7 |
| **200-ms latency** | 12.53 | 12.4 | 12.43 | 12.4 |

Table 19. Results comparing the simulation predictions (sim) and the actual measurements on the TV display (TV). We tested two conditions: where the ray traces were drawn from the eye CoR, (column CoR), and where the ray traces were drawn from the visual axis of the entrance pupil positions, after accounting for rotation due to vestibular ocular reflex and angle kappa (5°), (column Visual Axis). We measured 1) the magnitude of disparity when the head faces forward (0°) (Row Head forward), 2) the total translation on the screen for a monocular image with 0-ms latency during head rotation (Row 0-ms latency), and 3) the total translation on the screen for a monocular image with 200-ms latency during head rotation (Row 200-ms latency).

## 5 VALIDATING HIGH SENSITIVITY TO E2E LATENCY

In the AR experiment, a subset of subjects had detection thresholds that were 1 ms or less, including one participant who achieved 100% correct on all trials. We were motivated to quantify the magnitude of geometric error that is generated by 1 ms of latency to compared against known psychophysical thresholds. Using data for a representative participant whose detection threshold was 0.68 ms, we derived the disparity added by 0.68 ms of latency. We used the subject's IPD (61.397 mm) and head dynamics (Average maximum head rotation: 15.628° at 0.5 Hz). For these head motion dynamics we find that 1 ms of latency introduces a maximum error of 29 arcseconds 12. This falls within the range of previous findings of acuity from vision science psychophysical literature [Legge and Campbell 1981; Mckee and Nakayama 1984; Stevenson et al. 1989].

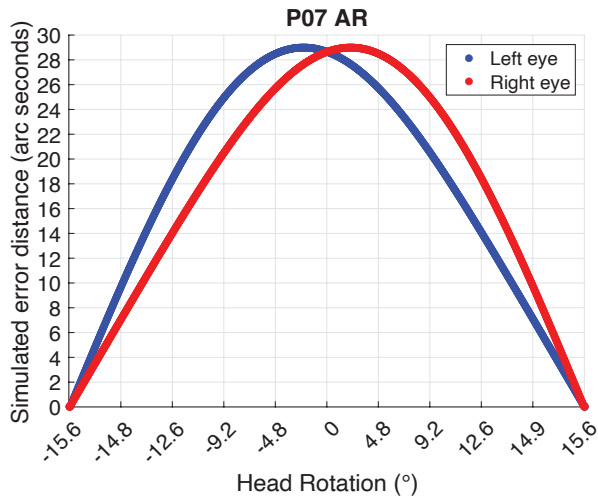

Fig. 12. Geometric error with 1 ms of latency for one participant. We elected to use a head rotation speed at the lower end of their head motion trajectories to be conservative.